\documentclass[journal]{IEEEtran}

\ifCLASSINFOpdf
\else
\fi

\usepackage{url}
\usepackage[nolist]{acronym}
\usepackage{amsfonts}
\usepackage{xcolor}
\usepackage{amsmath}
\usepackage{graphicx}
\usepackage[normalem]{ulem}
\usepackage{footnote} 
\usepackage{subfigure} 
\usepackage{bm}

\usepackage{amsmath}

\usepackage{url}
\usepackage{multirow}
\usepackage{amssymb}
\usepackage{tabularx} % in the preamble
\usepackage{booktabs}

\begin{document}

\newacro{WER}[WER]{Word Error Rate}
\newacro{CER}[CER]{Character Error Rate}
\newacro{AVSR}[AVSR]{Audio-Visual Speech Recognition}
\newacro{ASR}[ASR]{Automatic Speech Recognition}
\newacro{CTC}[CTC]{Connectionist Temporal Classification}
\newacro{CE}[CE]{Cross Entropy}
\newacro{RNN-T}[RNN-T]{Recurrent Neural Network Transducer}
\newacro{SNR}[SNR]{Signal-to-noise ratio}
\newacro{AVSE}[AVSE]{Audio-Visual Speech Enhancement}
\newacro{MFCC}[MFCC]{Mel-Frequency Cepstral Coefficients}
\newacro{HMM}[HMM]{Hidden Markov Models}
\newacro{STFT}[STFT]{Short-time Fourier transform}
\newacro{MFCC}[MFCC]{Mel-Frequency Cepstral Coefficients}
\newacro{CNN}[CNN]{Convolutional Neural Network}
\newacro{LSTM}[LSTM]{Long Short-Term Memory}
\newacro{GRU}[GRU]{Gated Recurrent Unit}
\newacro{DNN}[DNN]{Deep Neural Network}
\newacro{HMM}[HMM]{Hidden Markov Model}
\newacro{RNN}[RNN]{Recurrent Neural Network}
\newacro{STOA}[SOTA]{State-of-the-Art}
\newacro{FPS}[FPS]{Frames per Second}
\newacro{FFN}[FFN]{Feed Forward Network}
\newacro{MHSA}[MHSA]{Multi-Head Self Attention}
\newacro{KL}[KL]{Kullback-Leibler}
\newacro{MTL}[MTL]{MultiTask Learning}
\newacro{AV-VAD}[AV-VAD]{Audio-Visual Voice Activity Detection}
\newacro{RBM}[RBM]{Restricted Boltzmann Machines}
\newacro{LM}[LM]{Language Model}
\newacro{PUnet}[P\&U net]{Predict-and-Update Network}
\newacro{DL}[DL]{Deep Learning}
\newacro{FE}[FE]{Factorize-Excitation}

% \newacro{GRU}[GRU]{Gated Recurrent Unit}

\newcommand{\todo}[1]{\color{green} #1}
\newcommand{\etal}{\textit{et al.}}
\newcommand{\eg}{\textit{e.g.,}~}
\newcommand{\etc}{\textit{e.t.c.}}
\newcommand{\ie}{\textit{i.e.,}~}
\newcommand{\algs}{\textit{Algorithm}}

\newcommand{\fig}[1]{Fig.~}
\newcommand{\tab}[1]{Tab.~}
\newcommand{\Sec}[1]{Sec. }
\newcommand{\eq}[1]{Eq.}

\newcommand{\tofill}[1]{{\color{red}#1}}
\newcommand{\checkhere}[1]{{\color{green}#1}}
\newcommand{\qian}[1]{{\color{magenta}#1}}
\newcommand{\mat}[1]{{\textit{\textbf{}}#1}}
\renewcommand{\thefootnote}{}

\definecolor{burntumber}{rgb}{0.54, 0.2, 0.14}

\title{Predict-and-Update Network: Audio-Visual Speech Recognition Inspired by Human Speech Perception}
% \title{Predict-and-Update Network: A Novel Framework of Early Fusion for Audio-Visual Speech Recognition }


\author{
Jiadong Wang,
\IEEEauthorblockN{ Xinyuan Qian,
~\IEEEmembership{Member,~IEEE},
Haizhou Li,~\IEEEmembership{Fellow,~IEEE}, 
% Montgomery Scott\IEEEauthorrefmark{3}, and
% Eldon Tyrell\IEEEauthorrefmark{4},~\IEEEmembership{Fellow,~IEEE}
}

%\IEEEauthorblockA{Department of Electrical and Computer Engineering, National University of Singapore, Singapore}
% \IEEEauthorblockA{\IEEEauthorrefmark{2}Twentieth Century Fox, Springfield, USA}
% \IEEEauthorblockA{\IEEEauthorrefmark{3}Starfleet Academy, San Francisco, CA 96678 USA}
% \IEEEauthorblockA{\IEEEauthorrefmark{4}Tyrell Inc., 123 Replicant Street, Los Angeles, CA 90210 USA}% <-this % stops an unwanted space
%  \thanks{ $\star$  indicates equal contribution}
}

\markboth{Journal of \LaTeX\ Class Files,~Vol.~14, No.~8, August~2022}%
{Shell \MakeLowercase{\textit{et al.}}: Bare Demo of IEEEtran.cls for IEEE Transactions on Magnetics Journals}



% for Transactions on Magnetics papers, we must declare the abstract and
% index terms PRIOR to the title within the \IEEEtitleabstractindextext
% IEEEtran command as these need to go into the title area created by
\IEEEtitleabstractindextext{%
\begin{abstract}
Audio and visual signals complement each other in human speech perception, so do they in speech recognition. The visual hint is less evident than the acoustic hint, but more robust in a complex acoustic environment, as far as speech perception is concerned. 
It remains a challenge how we effectively exploit the interaction between audio and visual signals for automatic speech recognition. There have been studies to exploit visual signals as \textit{redundant} or \textit{complementary} information to audio input in a synchronous manner. Human studies suggest that visual signal primes the listener in advance 
as to when and on which frequency to attend to. We propose a \ac{PUnet}, to simulate such a visual cueing mechanism for \ac{AVSR}.  
In particular, we first predict the character posteriors of the spoken words, i.e. the visual embedding, based on the visual signals. The audio signal is then conditioned on the visual embedding via a novel cross-modal Conformer, that updates the character posteriors.
We validate the effectiveness of the visual cueing mechanism through extensive experiments. 
% The proposed P\&U net outperforms the state-of-the-art \ac{AVSR} method with a \ac{WER} reduction of 16.3\% and 42.0\% under clean and noisy conditions on LRS2-BBC, respectively.
The proposed P\&U net outperforms the state-of-the-art \ac{AVSR} methods on both LRS2-BBC and LRS3-BBC datasets, with the relative reduced \ac{WER}s exceeding 10\% and 40\%  under clean and noisy conditions, respectively. 
%The results s show that the earlier the fusion takes place, the better speech recognition performance can be achieved, especially in noisy conditions.
% And cueing effect promotes perception in noisy environment .
\end{abstract}

}

\maketitle

% not be used in maketitle, but will appear (\ie, to be "transported")
\IEEEdisplaynontitleabstractindextext

% \fi
\IEEEpeerreviewmaketitle

\section{Introduction}
\label{sec:intro}

\IEEEPARstart{H}{umans} have developed five senses: smell, taste, balance, vision, and hearing as a result of  evolution. 
Among these senses, vision and hearing are primarily involved during social interaction and for effective perception. Human speech perception benefits from combining audio and visual modalities with their unique and complementary characteristics.

It is apparent that multi-modal solutions, \eg audio and visual, outperform their unimodal counterparts in speech processing tasks such as 
speech extraction \cite{pan2022selective}, active speaker detection \cite{tao2021someone}, and emotion recognition \cite{kim2013deep}.
%\textcolor{red}{analyze characteristic of audio and visual} 
\ac{ASR} performance deteriorates in the presence of acoustic noise~\cite{afouras2018deep, yadav2021pitch}, and so does human speech recognition: the pair of /m/ and /n/, /b/ and /d/ are acoustically less distinguishable under noise~\cite{massaro1998speech}. Visual signals become very useful in a noisy environment because they are not affected by acoustic noise. One of the challenges is that many-to-one phoneme-to-viseme mapping exists \cite{cappelletta2012phoneme}. In other words, multiple phonemes (up to 13) could be rendered with the same lip movement \cite{cappelletta2012phoneme}. The question is how to effectively make use of such an inexact visual cue in speech recognition.
% \begin{figure}[t]
%     \centering
%     \includegraphics[width=\columnwidth]{Figure/ASR.pdf}
%     \caption{Three types of fusion methods: a) Late fusion that fuses decisions of different modalities; b) Intermediate fusion that intakes context representations of modalities after uni-modal sequence modelling and c) Early fusion that merges local features of modalities before uni-modal sequence modelling, where a local feature are calculated from a very small segment of raw data instead of whole data, such as \ac{STFT} features.}
%     \label{fig:fusiontype}
% \end{figure}
\begin{figure}[t]
    \centering
    \includegraphics[width=\columnwidth]{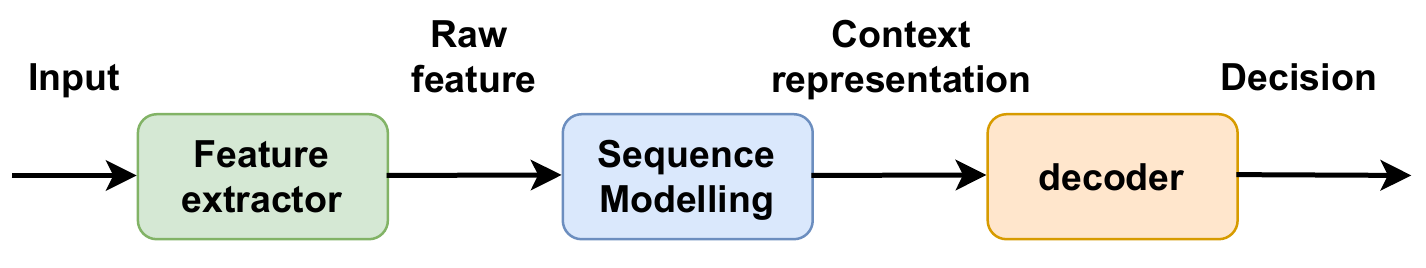}
    \caption{The universal diagram of end-to-end lip reading or \ac{ASR}. Input can be video or audio.}
    \label{fig:srdigram}
\end{figure}

In a human speech perception study~\cite{summerfeld1976some}, Summerfield made three hypotheses about the possible roles of video modality in improving noisy speech intelligibility. 1) Lipreading offers segmental and suprasegmental hints \textit{redundant} to audio hints. 
% 1) Lipreading offers segmental such as consonant and vowel, and suprasegmental, such as rhythm, stress, and intonation hints \textit{redundant} to audio hints. 
In other words, the phonetic information of speech is equally available in both audio and video modalities. For example, consonants and vowels are segmental, while rhythm, stress, and intonation are suprasegmental. 2) Lipreading offers segmental and suprasegmental hints \textit{complementary} to audio hints. That is, visual hints provide information unavailable in audio hints, perhaps due to acoustic interference. 3) When one listens, the audio signals (attended speech) and the video signals (lip movements) share common spatial-temporal properties, which may guide listeners to focus on speech signals of interest rather than acoustic interference.

% The hypotheses of the first two roles of video modality in human speech perception, namely providing hints \textit{complementary} or \textit{redundant} to audio hints, have motivated the implementation of automatic speech recognition. Visual hints can be decision-level  \cite{luettin2001asynchronous, stewart2013robust} or feature-level \cite{sterpu2020teach, lee2020audio}. If visual hints is decision-level, we infer decisions of individual modalities separately and fuse decisions via some str. For instance of decisions fusion, decisions can be integrated by weighted summation \cite{luettin2001asynchronous, stewart2013robust} or network \cite{yu2021fusing}.
% Intermediate fusion is the most adopted strategy. It conducts sequence modelling on audio and visual signals individually to generate context representations of individual modalities, and merge them later, 
% before making final decisions. It is also referred to as representation fusion where representations can merge via concatenation with \cite{sterpu2020teach, lee2020audio} or without \cite{ma2021end} audio-visual alignment.
% Early fusion seeks to integrate audio and visual features to an audio-visual feature at the early stage, namely, before sequence modelling of individual modalities. Afterwards, sequence modelling and decision making are conducted based on the audio-visual feature. To integrate audio and visual features, feature concatenation is applied~\cite{serdyuk2022transformer, makino2019recurrent, shi2022learning}.

The studies of the first two hypotheses in human speech perception serve as the motivations for implementing of  \ac{AVSR}. As audio and visual signals are \textit{redundant} or \textit{complementary}, they can be fused at the decision level or feature level for improved robustness.
At the decision level, we may fuse the decisions made by individual modalities through weighted summation \cite{luettin2001asynchronous, stewart2013robust} or network \cite{yu2021fusing}. % \textit{redundant} or \textit{complementary} decisions, \ac{AVSR} networks of this type are robuster than \ac{ASR} networks.
At the feature level, the fusion of either raw features or context representations was studied. 
A straightforward solution is to first combine raw audio and visual features via concatenation~\cite{serdyuk2022transformer, makino2019recurrent, shi2022learning}  and then apply sequence modeling and decoding. On the other hand, we may first encode the context representations of audio and visual signals from the short-time (raw) features, then fuse them, for example, {by concatenation} with audio-visual alignment~\cite{sterpu2020teach, lee2020audio} or without~\cite{ma2021end}.

In the third hypothesis, a visual cueing mechanism is suggested: lip movement preceding voice \cite{golumbic2013visual} can influence hearing at an elementary level, cueing listeners when and on which frequency of speeches one should focus \cite{grant2000use, grant2001effect}. {As well known, plosives consist of closure and release in order where only the latter is audible while both are visible \cite{byrd199354}. Thus, visible closure may help predict the coming voice \cite{golumbic2013visual}.} In a psycho-acoustic experiment, human subjects attempted to distinguish 10 different French syllables pronounced with similar lip movements ~\cite{schwartz2004seeing}. Lips were found to move before the arrival of the voice, which prompts listeners to pay attention to sounds, resulting in improved intelligibility. %\textcolor{blue}{Therefore, they attribute the improvement to audio cued by the preceding video.}
A similar visual cueing effect was reported in the experiment on bird song identification \cite{varghese2012visual}. It was shown that if humans know when a bird song is playing, they can better identify the specific bird to which the bird song belongs in the acoustic interference without any additional cue, such as a spatial cue. In this case, the simple indication of the onset and offset of the target song in the interference benefits human perception. 

Furthermore, in the study of the video modality that helps human speech detection, the high similarity between lip-opening area functions and acoustic envelope bands ($F2, F3$) is found~\cite{grant2000use} {with up to 165 ms of audio lag}, which shows the potential of video to reduce temporal and frequency uncertainty. Further experiments about human speech detection demonstrate that, with video, results of F2-filtered speeches by bandpass are comparable to those of unfiltered speeches, while those of F1-filtered speeches are not comparable \cite{grant2001effect}. Therefore, preceding the corresponding voice, video modality cues listeners to focus when in time and at which frequency on the spectrum.

% \cite{yu2021large} implements feature concatenation (a representative early fusion) on uncontrolled dataset. Although variations of this method  \cite{ibrahim2015feature, rahmani2018audio} show significant improvement to audio-only system with controlled utterances (digits) on noisy environment, there is not significant improvement (13.7\% relative) on uncontrolled dataset between audio-only model and early fusion \ac{AVSR} even in 0 dB noise \cite{yu2021large}, which can be 46.9\% with intermediate fusion \cite{afouras2018deep}. 
%  What limits early fusion may be that features of different modalities are inherently different \cite{michelsanti2021overview} and uncontrolled phrases aggravate this advantage. Thus, proper method for feature normalisation, transformation need more development \cite{michelsanti2021overview}. Besides, early fusion is hard to adopt pre-trained unimodal models as late fusion \cite{michelsanti2021overview}. Thus, an early-fusion framework, which is applied on the scenario of uncontrolled words and addresses the above two problem, is desirable. 

\begin{figure}[]
    \centering
    \includegraphics[scale=1]{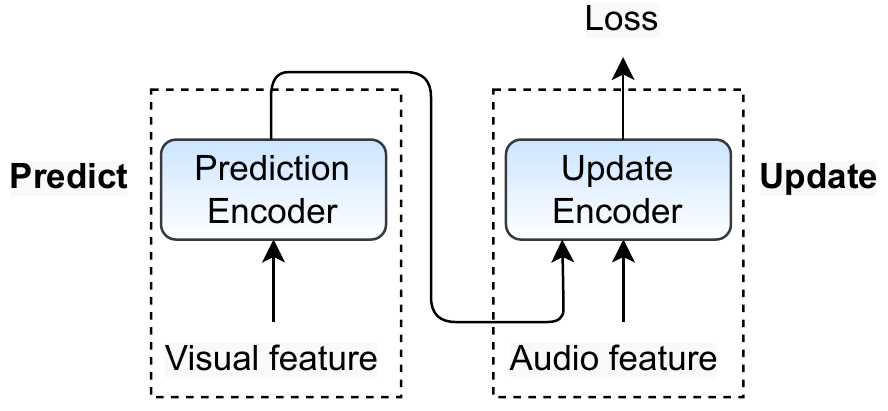}
    \caption{The general block diagram of our proposed \ac{PUnet} which adopts video modality to predict a sequence of a coarse probability distribution, \ie visual embedding, towards the target text transcription. The audio input is then augmented with the visual embedding for speech recognition via an update encoder.}
    \label{fig:framework}
\end{figure}

%\textcolor{red}{associate particle filter with avsr from their similarity of input modality}

We are motivated to implement the mechanism in the third hypothesis in a neural architecture. The question is how to effectively benefit from the visual cue with computational efficiency. We propose to use the prediction of spoken characters from the visual signal as the visual cue.
In view of the fact that video modality is ambiguous but noise-invariant while audio modality is explicit but noise-vulnerable, we adopt a
predict-and-update network architecture, where particle filter \cite{doucet2000sequential, kalman1960new} and Kalman filter \cite{kalman1960new} are employed for a pair of sensory inputs in robotics, just like audio and visual signals. %s whose characteristic is similar with the pair of audio and video modality in \ac{AVSR}. 
%Taking the task of orientation estimation as an instance, the estimation by measurements of a gyroscope alone leads to drift over time because it can only measure the angular velocity. However, the noise in the measurements is fairly small.  On the other hand, the measurements of the accelerometer are highly related to the orientation if there is no noise~\cite{kok2017using}.
%The aforesaid two filters first transform the gyroscope's measurements to target orientation as the prediction and then take accelerometer's measurements and the prediction to update orientation estimation. 

The predict-and-update network, \ie the \ac{PUnet}, emulates the visual cueing mechanism in human speech perception by considering the distinct characteristics of the audio and video modalities.
% We propose a \ac{PUnet} to a novel visual cueing audio mechanism to fuse distinct characteristic of audio and video modalities.
% to achieving fusion of visual cueing audio manner, while distinct characteristic of audio and video modalities are taken into account. 
{As shown in \fig~\ref{fig:framework}, we first take visual input to predict the probability distributions of spoken text in terms of characters, which is called \textit{visual embedding}. We then fuse the visual embedding with the audio features to update the probability distributions.} %encoder intakes the visual prediction to cue local audio features and meanwhile update the distribution sequence.
% Although sequence modelling may be adopted in the prediction encoder, we regard the \ac{PUnet} as a generalized early fusion since local audio features involve.
The visual cueing mechanism in the update encoder can be implemented using a factorized-excitation \ac{FFN}  with the same parameter structure as vanilla \ac{FFN}. %and in the updating decoder. %It is worth mentioning that the vanilla \ac{FFN} can achieve this  thus facilitates the network to leverage the well-developed unimodal models. 
Our contributions are summarized as follows:
\begin{enumerate}
    \item We propose a \ac{PUnet} to emulate the visual cueing mechanism in human speech perception while considering the heterogeneous characteristics of audio and visual signals. 
    \item We design a factorized-excitation \ac{FFN} for audio sequence modelling that effectively benefits from the visual cues.  
%    \item 
    % We conduct comprehensive experiments in which the 
 %   Results demonstrate that the manner of visual cueing audio at an early stage benefits from both the cueing mechanism and cueing raw audio feature. This will encourage researchers to explore early fusion strategy and the cueing mechanism in other audio-visual tasks.
    % \item We conduct comprehensive experiments in which the results demonstrate that fusion in an earlier stage yields better performance. This will encourage researchers to explore the early fusion strategy in other audio-visual tasks.
    % \item Experimental results convey a meaningful message that unimodal sequence modelling will lose some information beneath local feature, encouraging researchers to explore early fusion in other audio-visual tasks.
    \item The proposed audio-visual speech recognition network outperforms other state-of-the-art \ac{AVSR} methods~\cite{ma2021end} by a large margin, especially in a noisy environment. 
\end{enumerate}
% which alleviates the second problems of early fusion.
%  For comparison
% we also show the orientation errors from dead-reckoning the gyroscope measurements in the bottom plot
% (see also §1.2). These errors can be seen to drift over time.
% gyroscopes measure the specific force and the angular
% velocity
% gyroscope bias is fairly sma

% The orientation estimates obtained from the
% accelerometer and the magnetometer measurements are noisy but accurate over long periods of time.
% In this work, 
% we mimic humans' audio-visual perception of visual cueing audio at early stage for \ac{AVSR} with the following contributions:

\section{Related Work}\label{sec:relatedwork}

We start by reviewing how audio-visual fusion is implemented in the literature to set the stage for this work.  The studies of the first two hypotheses in human speech perception have motivated many audio-visual fusion implementations. In particular, audio and visual signals are considered redundant, complementary, and synchronous.   

With decision-level fusion, we make the final decisions by combining the decisions of individual modalities~\cite{luettin2001asynchronous,stewart2013robust, abdelaziz2015learning}, where audio-visual alignment information is not fully exploited. For instance, one can multiply the audio and visual decision probabilities to make a final decision~\cite{massaro1998speech}. Note that audio and video modality have own unique properties, their contributions should be considered separately.
% to the final decision should be taken in perspective.
 In~\cite{luettin2001asynchronous}, reliability is implemented in making the final decision by weighting
% \qian{reliability is measured to affect the final decision by weighting}
between the audio (0.7) and visual (0.3) modality, as  audio is more representative in a clean environment. To account for a varying environment, dynamic weights of audio and video modality are applied in~\cite{stewart2013robust, meutzner2017improving}. In \cite{meutzner2017improving}, dynamic weights are controlled by a logistic function of estimated \ac{SNR}. Recently, decision probabilities are fused with their quality indicators of audio and video modalities (\ie \ac{SNR} for audio and facial action unit for video), for example, by a \ac{LSTM} or fully-connected layer.
The decision-level fusion techniques mostly combine text posteriors from two modalities for speech recognition. They follow the idea that audio-visual information is \textit{redundant} or \textit{complementary}. However, they do not explicitly use synchronization or interaction between audio and video~\cite{katsaggelos2015audiovisual}. %, let alone the visual cueing mechanism.

It is also straightforward to fuse audio and visual signals at the feature level. Fusion can take place between raw features \cite{serdyuk2022transformer, makino2019recurrent}, context representations~\cite{afouras2018deep, petridis2018audio,yu2020audio}, or a mix of both~\cite{liu2021audio, hu2019dense}. The fusion of context representations is widely used by concatenation~\cite{afouras2018deep, petridis2018audio} or multiplication~\cite{yu2020audio}. The feature fusion techniques seek to make use of the synchronization information. %, they treat audio and visual representations in the same way without considering their distinct properties~\cite{michelsanti2021overview}. %\textcolor{red}{to Jiadong: the following sentence was mentioned in the previous paragraph. we shouldn't mention the same thing in two different places - i understand that they are in different tasks. but this statement doesn't address the point that audio-visual have distinct properties} To address this problem, In \cite{zhou2019modality}, the reliability of the modalities is taken into consideration where a weighted sum of audio and visual context representations is employed with a dynamic weight estimator. 
{Some looked into audio-visual alignment before the feature concatenation by varying the hop size of short-time Fourier transform (STFT) \cite{makino2019recurrent} or resampling visual signals to the rate of the audio spectrum \cite{serdyuk2022transformer}. Others studied the audio-visual interaction by} cross-modal attention, where audio context representation is used to query visual context representation to generate audio-aligned visual representation~\cite{sterpu2020teach, paraskevopoulos2020multiresolution}. This mechanism offers a temporally aligned visual representation to the audio representation. Similarly, visual context representation can also be used to query audio context representation~\cite{lee2020audio} to fully use the audio-visual interaction. %Moreover it increases the role of video modality, given the observation that audio is unequivocal under clean conditions to de-emphasize vision's influence. 
%To increase the contribution by the visual signals, it was studied to use audio and visual sig  \cite{lee2020audio} apply both AV align and VA align (use visual context representation to query audio context representation). 
%When fusion happens between raw features, concatenation is adopted in~\cite{afouras2018deep, petridis2018audio,yu2020audio}. 
There was also an attempt~\cite{ngiam2011multimodal} to map audio and visual raw features to a shared representation to normalize the inherent different modalities. In general, feature-level fusion techniques seek to exploit the audio-visual synchronization property. They have not exploited the asynchronization between audio and visual signals at the time of feature fusion, where the visual signal is processed ahead of the audio signal.

In general, all fusion studies are motivated by the belief that audio and visual signals are redundant and complementary in a synchronous manner. They have not made use of the audio-visual interaction as discovered in the third hypothesis~\cite{summerfeld1976some} by Summerfield. We are motivated to explore such audio-visual interaction in this work.

\section{Predict-and-Update Network}\label{sec:proposal}
We now propose a novel neural architecture, \ac{PUnet}, for \ac{AVSR}, as shown in \fig~\ref{fig:architecture}. In particular, we will elaborate on how the predict-and-update framework emulates the human visual cueing mechanism.
%We start by the feature extractors of both modalities. We then separately present procedures of the prediction and update step,
% \qian{especially  a cross-modal Conformer block with a novel factorized-excitation \ac{FFN} module designed for audio-visual fusion through the cueing mechanism}
%especially a cross-modal Conformer block with a factorized-excitation \ac{FFN} module which is the implementation of cueing mechanism. Finally, we describe the training criteria.

\subsection{Audio and visual features}
\label{audio_feat}
Spectral features are widely used as acoustic features for speech signals in speech recognition~\cite{afouras2018deep, lee2020audio}. % \cite{maurer2000formant}, frequency-based features are widely adopted for speech recognition\cite{afouras2018deep, lee2020audio}. 
%As video modality cues listeners when and on which frequency of speeches to focus, we would like to choose a temporal-frequency audio feature. Therefore, 
We adopt \ac{STFT} as the spectral feature, as in other \ac{AVSR} studies~\cite{afouras2018deep, xu2020discriminative}. To compute the spectral features, we adopt a window size of 40 ms and a window shift of 10 ms %We compute the spectrum on speeches sampled at a sample rate of 16 kHz  with a 40 ms window and 10 ms hop-length, 
that converts a speech utterance into a sequence of 321-dimensional speech frames. Subsequently, a \ac{CNN} module is adopted to down-sample one in four frames so that the frame rate of speech features is aligned with that of the video frame sequence at 25 \ac{FPS}. 

%\subsection{Visual features}
%\label{visual_feat}
The visual signal is sampled at 25 \ac{FPS}. We crop a 122$\times$122 patch from each 224$\times$224 image   for pre-processing. 
%After the collection of large-scale \ac{AVSR} datasets \cite{afouras2018deep, makino2019recurrent}, 
%It is common that \ac{CNN} features gain more popularity for reading unconstrained sentences than hand-craft features \cite{zhou2014review}, such as VGG-M model \cite{chatfield2014return} and ResNet \cite{he2016deep}. 
We adopt ResNet~\cite{he2016deep} to generate visual features considering both performance and training effort~\cite{he2016deep}.
% VGG-M model \cite{chatfield2014return} is adopted in \cite{son2017lip} where five continuous frames of lip images are used to extract local features. Afterwards, to facilitate back-propagation in the deep network, 3D \ac{CNN} and ResNet \cite{he2016deep} are combined as feature extractors \cite{afouras2018deep,ren2021learning}. ResNet is preferable as it is easier to converge, and boosts performance when networks become deeper \cite{he2016deep}.}
Therefore, we apply a 3D convolutional layer followed by a 2D ResNet-18 \cite{he2016deep} on image sequences as the visual extractor. The kernel size of the 3D convolutional layer is 5$\times$7$\times$7 (time, height, width). After ResNet-18, we employ a global average pooling layer to reach the expected feature dimensions.

% \begin{figure}[]
%     \centering
%     \includegraphics[scale=0.5]{Figure/lip_aware.pdf}
%     \caption{End-to-end audio-visual speech recognition architecture}
%     \label{fig:architecture}
% \end{figure}

\subsection{Prediction}
\label{subsec:prediction}

\begin{figure*}[!tb]
\begin{center} 
\subfigure[]{ \label{subfig:conformer}
\includegraphics[height=5.5cm]{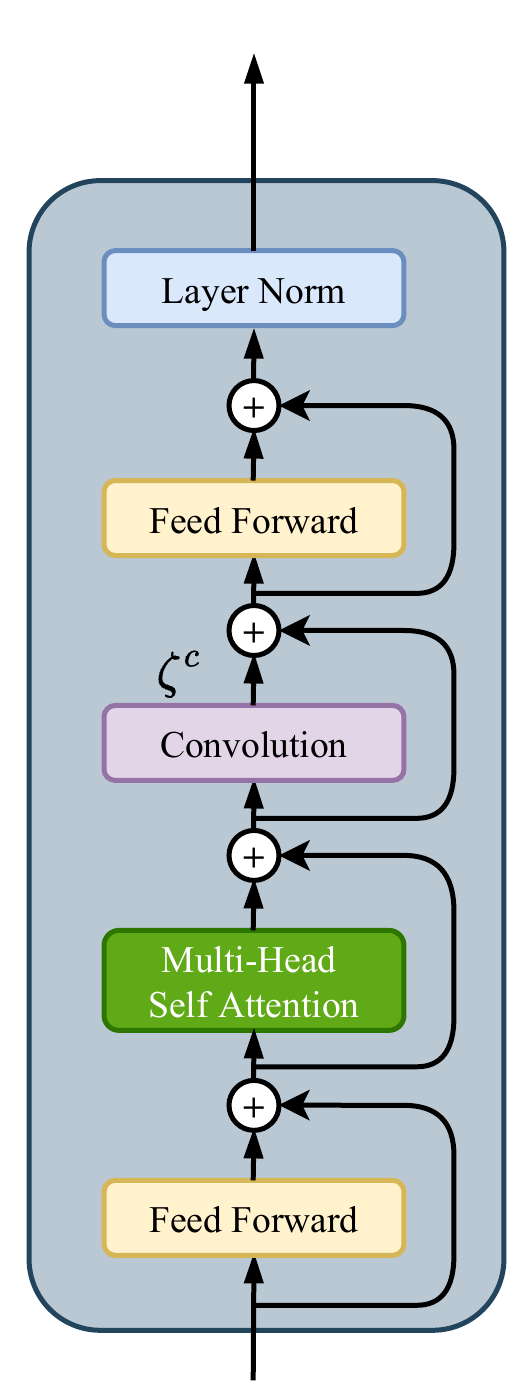}}
\subfigure[]{\label{subfig:ccconformer}
\includegraphics[height=5.5cm]{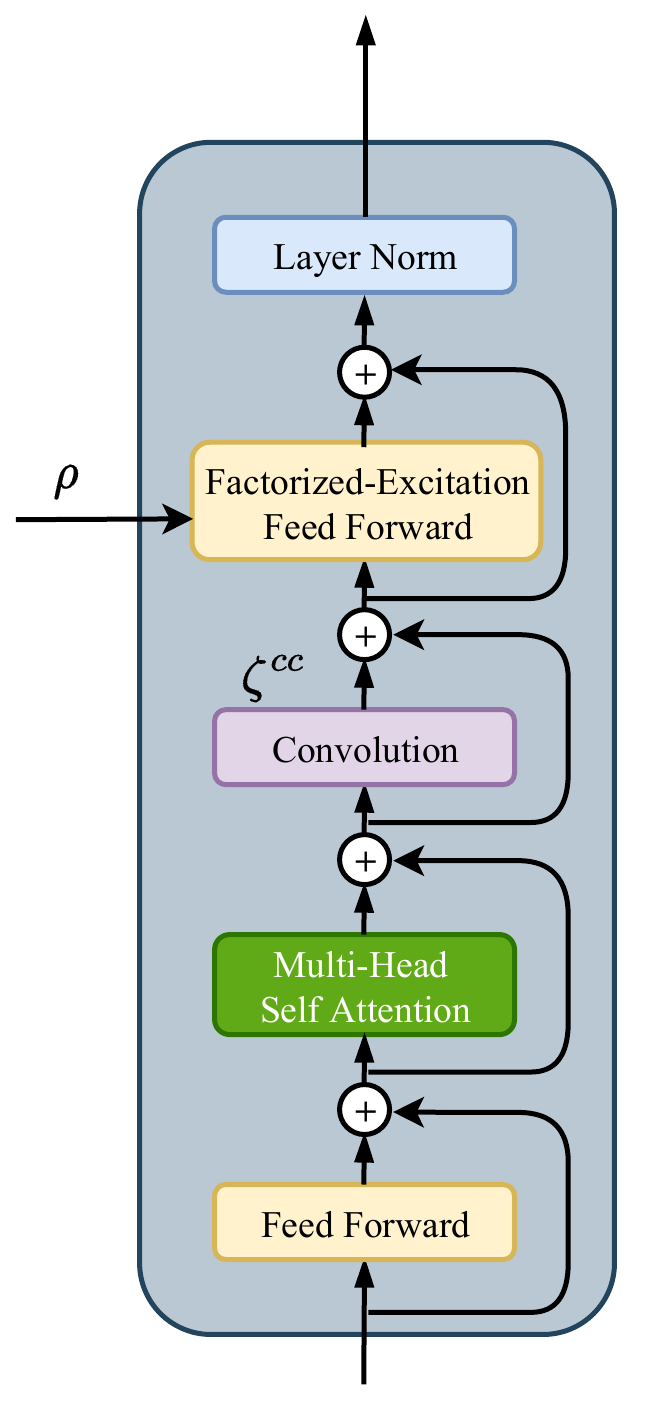}}
\subfigure[]{\label{subfig:punetearly}
\includegraphics[height=5.5cm]{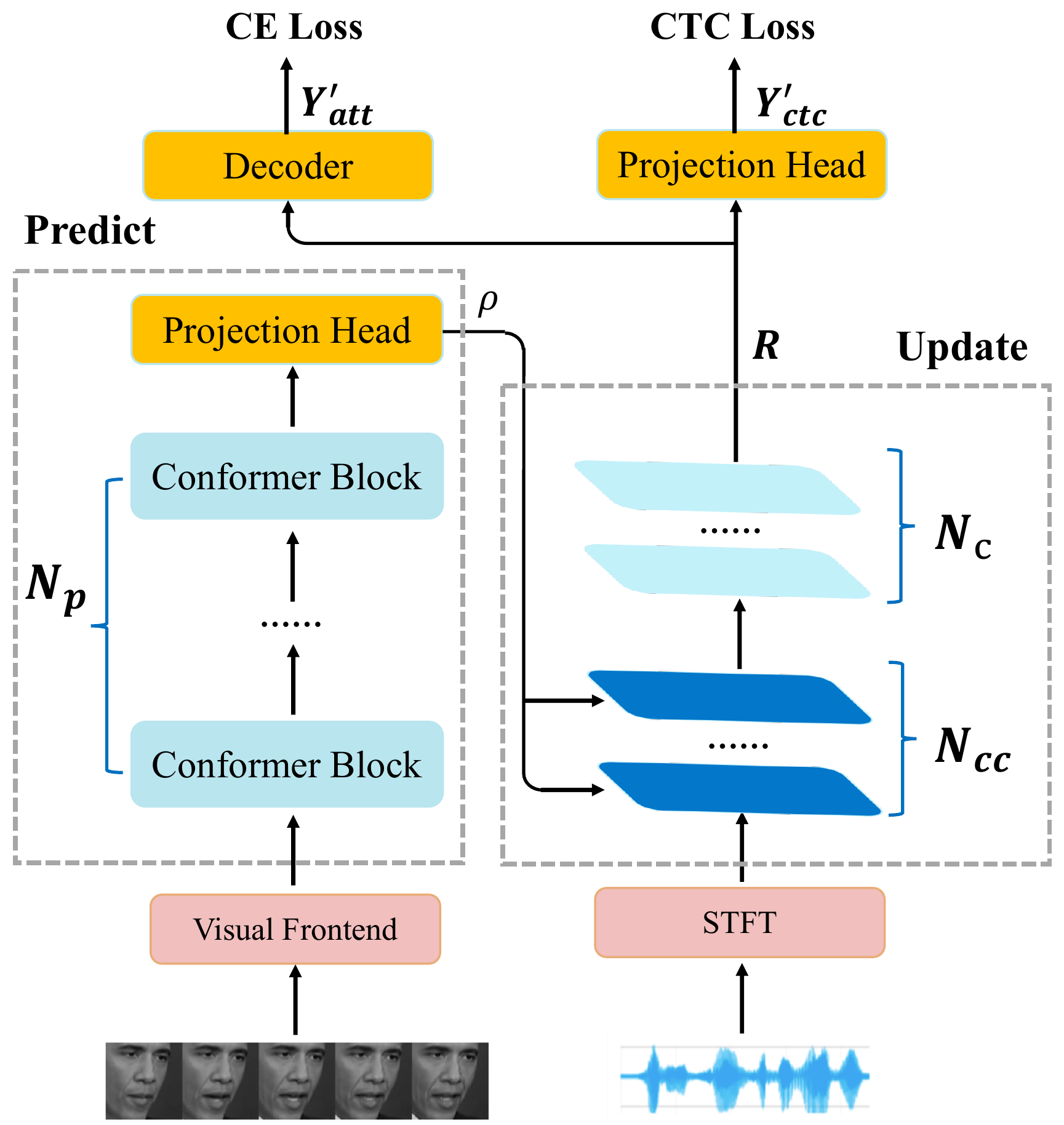}}
\subfigure[]{\label{subfig:punetlate}
\includegraphics[height=5.5cm]{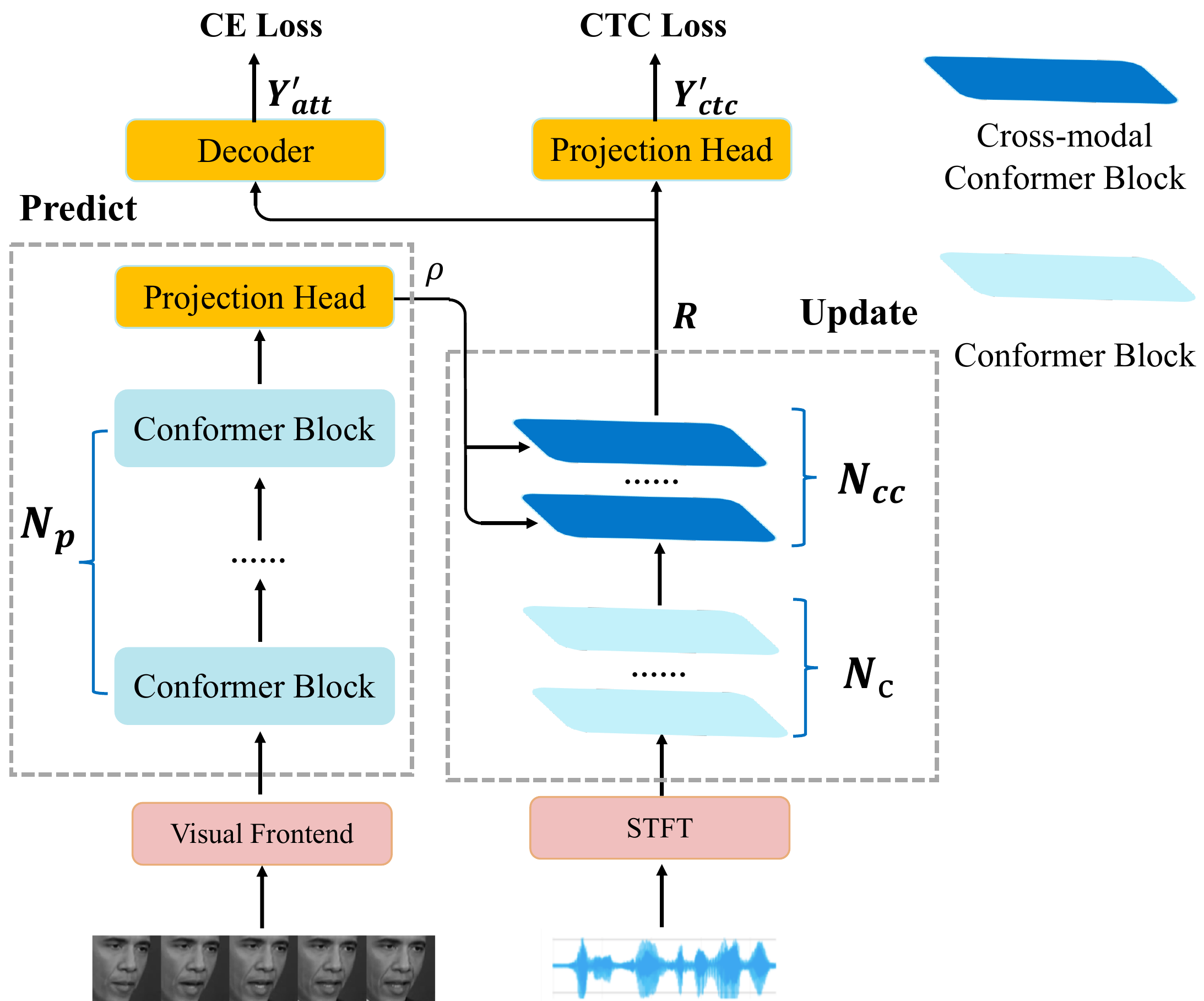}}
\end{center}
  \caption{Neural architecture of the predict-and-update network: a) Conformer block; b) Cross-modal conformer block. $\rho$ denotes 
%   the predicted character posteriors
 visual embedding from visual input; $\zeta^c$ and $\zeta^{cc}$ are the outputs of the convolutional blocks in the Conformer and the cross-modal Conformer, respectively and  c) {our proposed} \ac{PUnet} (early) where the audio input is conditioned on the visual embedding at an early stage, d) {a contrasive architecture of} P\&U net (late) where the audio input is conditioned at an late stage. Notes: $N_p$ is the number of the Conformer blocks in the prediction encoder; $N_c$ and $N_{cc}$ are the number of the Conformer and the cross-modal Conformer block in the update encoder.  {$Y'_{att}$ and $Y'_{ctc}$ are hypotheses about text sequence for \ac{CE} loss and \ac{CTC} loss, respectively. $R$ is the audio-visual context representation.}} 
%   \label{fig:platform}
  \label{fig:architecture}
\end{figure*}

We design a predictor to generate a contextual representation, \ie visual embedding, from a video sequence. The predictor predicts the arrival of sound ahead of time. To this end, an encoder is required to encode the temporal information from a video sequence. There are many options for encoders, such as \ac{LSTM} \cite{ sterpu2020teach, hochreiter1997long}, \ac{GRU} \cite{xu2020discriminative, cho2014learning}, and Transformer \cite{afouras2018deep, vaswani2017attention}. Despite great success, it was argued that these encoders 
have not explicitly leveraged short-range temporal dependencies~\cite{zhang2019spatio}. %Thus, \cite{zhang2019spatio} proposes a novel encoder with a \ac{CNN}-based temporal focal block and local self-attention. 
The fully-\ac{CNN} encoder is one of the solutions for \ac{ASR} \cite{li2019jasper} and lip-reading \cite{afouras2020asr}. Recently, Conformer \cite{ma2021end, gulati2020conformer} has shown outstanding performance taking advantage of both \ac{CNN} and the transformer, which are good at modelling local and global dependencies~\cite{ma2021end}. We therefore adopt Conformer as our encoders.  %Noteworthy, \cite{ma2021end} achieves state-of-the-art performance for \ac{AVSR} without external data.

As shown in \fig~\ref{subfig:punetearly}-\fig~ \ref{subfig:punetlate}, the prediction network consists of some stacked Conformer blocks \cite{gulati2020conformer} and a projection head. The Conformer blocks model temporal dependency in the visual sequence, while the projection head compresses the visual representation to a low-dimension character posterior with a softmax function. As shown in \fig~1, a Conformer block consists of a \ac{FFN} layer, a \ac{MHSA} module, a convolutional block, and another \ac{FFN} layer in a pipeline. A layer normalization module is applied before each sub-module in the Conformer block. 

The \ac{MHSA} module takes as input the query $Q\in \mathbb{R}^{T\times d_k}$, key $K\in \mathbb{R}^{T\times d_k}$ and value $V\in \mathbb{R}^{T\times d_k}$ with relative positional encoding, where $T$ is the number of frames in an utterance and $d_k$ is the dimension of query, key, and value. The \ac{MHSA} module first maps the set of three inputs to $h$ sets according to $h$ heads and then calculates the scaled dot-product attention which is depicted in \cite{vaswani2017attention}. The convolutional block consists of two pointwise 1D convolutional layers and a depthwise 1D convolutional layer with a skip connection. It outputs a hidden audio feature $\zeta^c \in \mathbb{R}^{T\times d_a}$, where $d_a$ is the feature dimension.
% the combination of two pointwise 1D convolution layer and a depthwise 1D convolution with a skip connection. 
By combining the convolutional block and the self-attention module \cite{vaswani2017attention}, the Conformer as an encoder is capable of modeling both local and global interactions of visual features. 
The \ac{FFN} process is formulated as:
\begin{equation}\label{eq:ffn}
 FFN(\tau) = W_2 \times \o(W_1\tau^{\intercal} +B_1)+B_2
\end{equation}
%where $\o$ denotes the ReLU activation function. 
where $\tau \in \mathbb{R}^{T\times d_a}$ is the input. $W_1 \in \mathbb{R}^{d_{ff}\times d_a}$, $B_1\in \mathbb{R}^{d_{ff}}$ and $W_2 \in \mathbb{R}^{d_a\times d_{ff}}$, $B_2\in \mathbb{R}^{d_a}$ are the parameters of the first and  second linear layers. $\o$ is an activation function. {In short, \ac{FFN} firstly transforms $\tau$ from $d_a$-dimensions to $d_{ff}$-dimensions and then transforms it back to $d_a$-dimensions}.

Finally, the projection head generates the visual embedding $\rho \in \mathbb{R}^{T \times C}$, which is a sequence of $T$ frames. Each frame is represented by a $C$-dimension character posterior. The prediction module can be pre-trained on a lipreading task, which aligns $\rho$ with the target character by \ac{CTC} and attention decoder loss to ensure that the predictor outputs are phonetically relevant to the output text. The training loss will be discussed in \Sec~\ref{subsec:loss}.

\subsection{Update}

%\textcolor{red}{to Jiadong: i think you have answered most of my questions. WHy do you revise the paragraph to reflect what you say about Update module. Please use the Prediction module as an example for writing. please don't just answer my question - please revise your text. i can help fine-tune. Your current text doesn't explain the design purpose and the function of Update module. You only explain how it looks like, which is not the most important.  We need to start a sentence such as `We design a update module to ...'. }

We design an update encoder in \fig~\ref{subfig:punetearly} to produce an audio-visual context representation from the visual embedding and the raw audio feature. This encoder emulates the visual cueing mechanism. Then the audio-visual context representation is passed to the decoder and the \ac{CTC} module.
% text posteriors.
% \textcolor{blue}{The encoder in the update step (\ie update encoder) undertakes two sequential roles: 1. to adopt the prior visual prediction to cue noise-vulnerable audio features focus on target-related elements, namely, to fuse audio and video modalities 2. to further update (transform) visual-cued audio features to a context representation which is closer to text.}

% Multi-modal fusion takes place in the update step, which adopts the visual prediction to cue audio features to compensate their noise-vulnerable characteristic and meanwhile updates the prediction by the representative audio modality.
% In The audio modality is used to update the prediction of text probability distribution sequences due to its representative but noise-vulnerable characteristic.
% To perform cueing mechanism in the update step, visual prediction and hierarchical audio features are integrated in a stack of $N_c$ Conformer blocks and $N_{cc}$ cross-modal Conformer blocks, as shown in \fig~\ref{fig:architecture}.
In the update encoder, the audio-visual fusion is achieved by stacking $N_{cc}$ cross-modal Conformer blocks. In different cross-modal Conformer blocks, the visual embedding joins the hierarchical audio features, as the visual embedding in the form of character posteriors is closer to the output text than the hierarchical audio features in shallow blocks. This fusion emulates the visual-cueing mechanism. Visual-cued audio features are then processed by $N_c$ vanilla Conformer blocks, that generate audio-visual contextual representation, \ie the updated text posteriors. 
% To perform cueing mechanism in the update step, 
% the visual prediction cues hierarchical audio features in each of the stacked $N_{cc}$ cross-modal Conformer blocks.
% , as shown in \fig~\ref{subfig:punetearly}.
% Then, the visual-cued audio features are feed into another stack of $N_c$ vanilla Conformer blocks to make final updates.
% Besides, a stack of $N_c$ Conformer blocks are applied to further update the distribution sequences. 
% Noteworthily, the order of Conformer blocks and cross-modal Conformer blocks is changeable as shown in \fig~\ref{subfig:punetearly} and \fig~\ref{subfig:punetlate}, which provides comparison to explore the role of early fusion.

\begin{figure}[]
    \centering
    \includegraphics[scale=0.5]{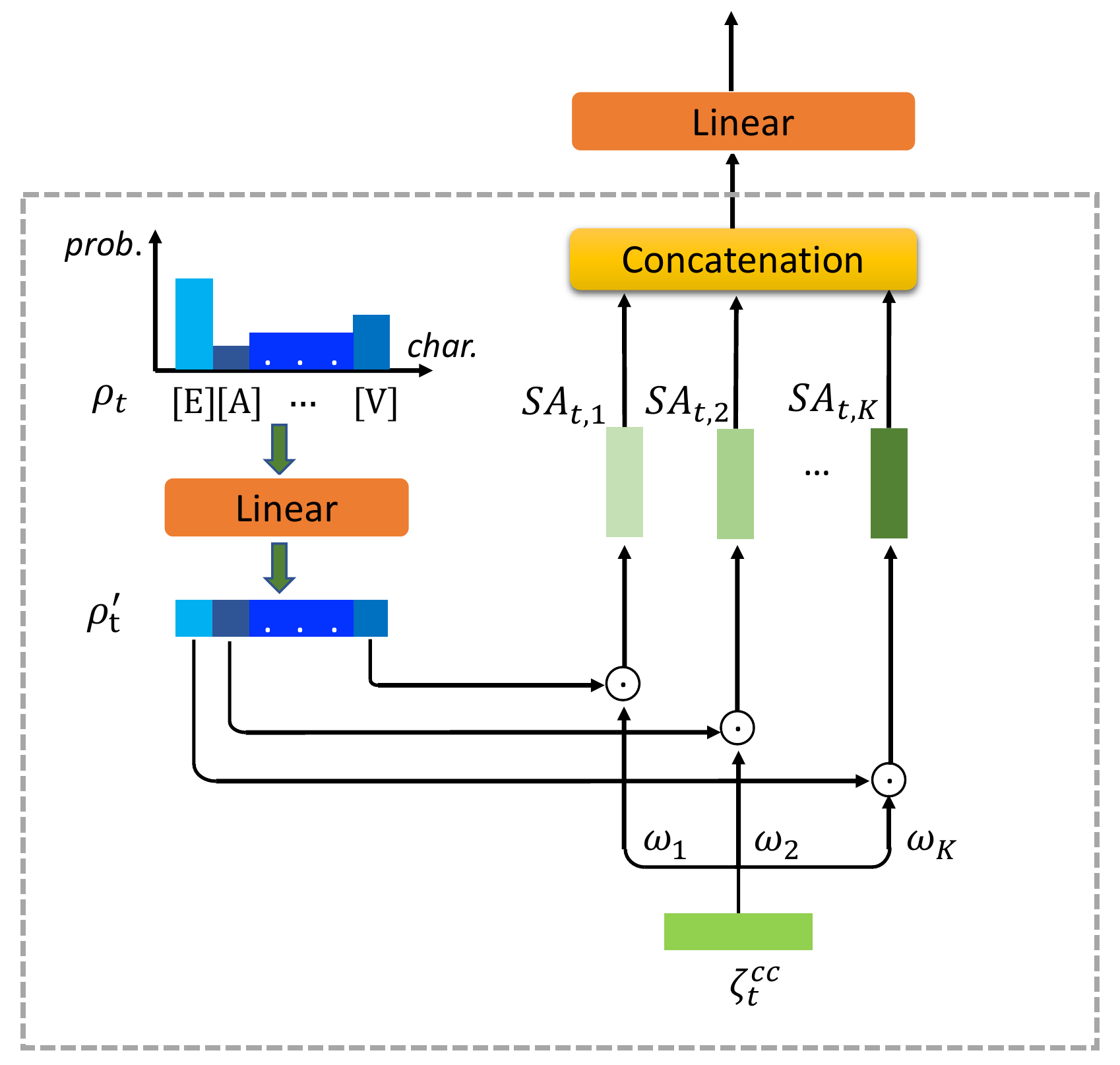}
    \caption{The diagram of the factorized-excitation \ac{FFN} module. An activation layer after the concatenation is omitted for brevity; $\zeta^{cc}_{t}$  and $\rho_t$ denote the hidden audio features and the 
    % predicted \textit{character posteriors} vector
    visual embedding from vision at frame $t$. $\omega_1$ ... $\omega_K$ are shared across all time frames.
    }
    \label{fig:factorized_feed_forward}
\end{figure}

The cross-modal Conformer block is a revised version of the vanilla configuration~\cite{gulati2020conformer} by replacing the original \ac{FFN} \cite{vaswani2017attention} after the convolutional block with a proposed factorized-excitation \ac{FFN}, as illustrated in \fig~\ref{fig:factorized_feed_forward}. We also propose a contrastive architecture by swapping the Conformer and the cross-modal Conformer blocks, as illustrated in \fig~\ref{subfig:punetlate}, to observe the effect between early fusion and late fusion.

%for multi-modal fusion. 
% \fig~\ref{fig:factorized_feed_forward} illustrates the factorized-excitation \ac{FFN} architecture in detail.
% The cross-modal Conformer block follows the design of the Conformer block \cite{gulati2020conformer}, excepting the \ac{FFN} after the convolution layer. A factorized-excitation \ac{FFN} substituting the original \ac{FFN} \cite{vaswani2017attention} practically performs modality fusion as shown in \fig~\ref{fig:factorized_feed_forward}. 

The factorized-excitation \ac{FFN} is inspired by the factorized layer, which has been implemented in \ac{ASR} \cite{delcroix2015context} and speech extraction \cite{vzmolikova2019speakerbeam}. It is typically used to condition the network on prior external knowledge. Thus, it is a perfect mechanism to take visual cues to strengthen speech encoding.  In \cite{delcroix2015context}, the factorized layer is used to adapt the \ac{ASR} model to acoustic conditions with varying speaker genders, identities and noise \ac{SNR}, where the results of different conditions are summed with external factors. The factorized layer is formulated as:

\begin{equation}\label{eq:factorized_layer}
 X^{i+1} = \sum_{k=1}^K \alpha_k(\omega_{k}X^i+b_k)
\end{equation}
where $K$ is the number of subspaces that represents the conditions. $X^i$ is an input of neurons in layer $i$ and is mapped to different subspaces through $\omega_k$ and $b_k$. $\alpha_k$ are external factors and can be calculated by clustering \cite{delcroix2015context}. 
In speech extraction~\cite{vzmolikova2019speakerbeam}, a reference from the target speaker's speech is used as a condition to guide speech extraction. Such a reference can be a fixed speaker embedding across time~\cite{dehak2010front}, or a frame-varying embedding sequence~\cite{gu2020multi}. The previous studies suggest that the factorized layer is effective in conditioning the network on prior knowledge for speech processing.  

% Therefore, the factorized layer is a good choice to achieve cueing mechanism. 
% fusion by directly applying the factorized layer at early stage
Although the factorized layer effectively employs prior knowledge, which is the visual embedding herein, how to design a cross-modal Conformer block with the factorized layer is worthy of discussion~\cite{joze2020mmtm}. We prefer a cross-modal Conformer block~\cite{gulati2020conformer} that shares the same parameter architecture as the vanilla Conformer block in the \ac{ASR} task, because such a cross-modal Conformer block could potentially be pre-trained on an abundantly available audio-only dataset.

Therefore, we design a factorized-excitation \ac{FFN} to modulate the audio features by conditioning on the visual embedding, and keep the same parameter architecture as the vanilla \ac{FFN}. 

The factorized-excitation \ac{FFN} replaces the first linear layer of the vanilla \ac{FFN}, as shown in the dotted grey box in \fig~\ref{fig:factorized_feed_forward}, and keeps the second linear layer of the vanilla \ac{FFN}. Therefore, we consider that the cross-modal Conformer preserves the vanilla Conformer architecture. The dotted box processes
% the output of the convolution block $\zeta^{cc}_{t}$ at \qian{frame t} as follows,
the convolutional block's output $\zeta^{cc}$ and the visual embedding $\rho$ frame-by-frame as follows, where we take the $t$-th frame as an example.
\begin{equation}\label{eq:factorized_ffn_0}
  \rho'_t = W_{\rho}\rho_t + B_{\rho}
\end{equation}
% where $\rho'_t\in \mathbb{R}^{K}$ is a variant of $\rho_t\in \mathbb{R}^{C}$,
where $\rho'_t\in \mathbb{R}^{K}$ is a variant of $\rho_t\in \mathbb{R}^{C}$,
% to avoid that $d_{ff}$ is not divisible by $C$,
$W_{\rho}$ and $B_{\rho}$ are parameters of the linear layer.

\begin{equation}\label{eq:factorized_ffn_1}
  SA_{t,k} = \rho'_{t,k}\omega_k \zeta^{cc}_{t} + b_k
\end{equation}
where $SA_{t,k} \in \mathbb{R}^{d_l} $ denotes the $k$-th subspace;
% at the $t$-th frame;
{$\zeta^{cc}_t$ is the output of the convolutional block taken by the factorized-excitation \ac{FFN} as the input},
% \textcolor{blue}{$t$-th frame of the convolution block's output $\zeta^{cc}_t$ is the $t$-th frame of the input to the factorized-excitation \ac{FFN}}
as shown in \fig~\ref{fig:architecture} and \fig~\ref{fig:factorized_feed_forward};
We transform $\zeta^{cc}_t$ to $SA_{t,k}$ by applying {$\omega_k \in \mathbb{R}^{d_l \times d_a}$ and $b_k \in \mathbb{R}^{d_l}$  with the prior knowledge $\rho'_{t,k}$. We make sure that $d_l \times K = d_{ff}$ so that the concatenation of all $w_k$ has the same dimension as $W_1$ in Eq.~\ref{eq:ffn}. So do all $b_k$. Thus, $\mathcal{F}_t$ has the same dimension as the output of the first linear layer of the vanilla \ac{FFN},}
% (concatenation of all $SA_{t,k}$)
% \textcolor{blue}{Finally, we concatenate all $SA_{t,k}$ and therefore the dimension of $\mathcal{F}_t$ is the same as the output of the first linear layer of the vanilla \ac{FFN} at time frame $t$.}
\begin{equation}\label{eq:factorized_ffn_2}
 \mathcal{F}_t = \o(Concat(SA_{t,1}, SA_{t,2}, ..., SA_{t,K}))
\end{equation}
where $\o$ is an activation function.
% $\mathcal{F}$ is hidden variables which is fed to the consequent linear layer. \textcolor{red}{(there is no need to define $\mathcal{F}$. )}

% $\rho'\in \mathbb{R}^{T \times K}$ is a variant of $\rho\in \mathbb{R}^{T \times C}$ by a linear layer to avoid that $d_{ff}$ is not divisible by $C$, where $W_{\rho}$ and $B_{\rho}$ is the parameters of the linear layer.

%\textcolor{burntumber}{To Prof. Li, because the factorized-excitation FFN replaces the first linear layer of vanilla FFN with dotted grey rectangle in \fig 4 while second linear layer remains, and we say our proposed factorized-excitation FFN does not change the parameter architecture. So I write the process of getting $\mathcal{F}$ (output of dotted grey rectangle) to explain why the parameter architecture does not change. I will revise the sentence to highlight it.}

With the factorized-excitation \ac{FFN}, the update encoder benefits from both the pre-trained acoustic model and the visual embedding.

\subsection{Loss function}
\label{subsec:loss}
We update the model parameters during training by gradient back-propagation subject to some loss functions. The model includes the mapping modules, the update encoder, the predictor, and the feature extractors. The mapping modules are used to predict text sequence from the audio-visual context representation.

An AVSR system~\cite{afouras2018deep, ma2019investigating} may employ the \ac{CTC} loss~\cite{graves2006connectionist}, %, which employs dynamic programming to ensure monotonic alignment
to force monotonic alignment between the audio-visual context representation and the text sequence, where it is assumed that audio-visual frames are independent of one another~\cite{watanabe2017hybrid}. 
The attention-based loss \cite{son2017lip} represents another line of thought, which considers inter-frame dependency via an attention mechanism.
% The attention module can help establish temporal alignment between the target text and the encoder outputs from the audio and video modalities.
However, the attention-based loss does not enforce monotonic alignment~\cite{petridis2018audio}. To benefit from the best of the two techniques, we adopt the hybrid \ac{CTC} / attention method~\cite{gao2021tran, ma2021end} as the loss function of the P\&U net, that delivers promising results~\cite{ma2021end}.

Let us denote $Y\in\mathbb{R}^{L \times C}$ as the ground truth, \ie the text sequence and $R\in\mathbb{R}^{T \times d_a}$ as the audio-visual context representation, where $L$, $T$, $d_a$ are the length of the text sequence, the number of frames of the audio-visual context representation, and the dimension of the update encoder of the audio-visual context representation, respectively. 

% Thus, we adopt hybrid \ac{CTC}/attention as the loss function of the P\&U net. Let us denote $Y=[y_1, y_2, ..., y_L]$ as the text sequence and $R=[r_1, r_2, ..., r_T]$ as the audio-visual context representation , \ie the output of the update encoder. We aim to find appropriate losses to align $R\in\mathbb{R}^{T \times d_a}$ with $Y\in\mathbb{R}^{L \times C}$, where $L$, $T$, $d_a$ are the length of the text sequence, the frame number, and output dimension of the update encoder, respectively. 
% As mentioned in \sec~\ref{sec:relatedwork}, hybrid \ac{CTC}/attention shows great performance because one component's merits can compensate for the defect of another.
% intakes the audio-visual context representation 
In practice, the mapping module of the \ac{CTC} loss is a linear layer followed by a softmax layer to get the posterior probabilities of $C$ classes at all frames $Y'_{ctc} \in \mathbb{R}^{T \times C}$. Afterwards, dynamic programming is used to align the text sequence $Y$ and posterior probabilities $Y'_{ctc}$.

% \textcolor{red}{(to jiadong: what is R Y? ------ you must say the ground truth sequence Y, the output of the update encoder R .... and you cannot just say R and Y.)}
% \textcolor{burntumber}{R and Y is defined in the first paragraph of this section, R is output of update encoder, Y is the ground truth. \textcolor{red}{ please let Xinyuan to revise this part, and I will come back to this paper once it is done. 1/ the symbols need to be useful, and consistent with the figures. there are too many inconsistency. 2/ we only define symbols when necessary for example to better explain the concept or to be used later on. If we can use figures and text to explain, we don't need to define the symbols.)}}

%\textcolor{red}{(to Jiadong/Xinyuan: is the output of the update encoder also a text distribution?))}

%\textcolor{burntumber}{To Prof. Li, I think it's similar, it is a text representation, as it will be fed to CTC loss. But it need a linear layer to transform it to text distribution. In experiments, $d_a$ is 256 while $C$ is 40.}

% hypothesis $Y'$

The mapping module of the attention loss is a transformer decoder \cite{vaswani2017attention, gao2022genre} consisting of several transformer blocks. During training, the transformer decoder is adopted to parallelly predict all texts in an utterance from an audio-visual context representation $R$ and a text sequence $Y$. All predicted texts form the text hypothesis $Y'_{att}$. The \ac{CE} loss is applied to reduce the error between $Y'_{att}$ and $Y$. During inference, the decoder generates characters one by one to form a character sequence. Unlike the Conformer, the transformer decoder uses absolute positional encoding.   
% For attention loss, a transformer decoder \cite{vaswani2017attention} which contains a positional encoding module and a set of multi-head attention blocks is employed. In each block, there are a masked multi-head self-attention, a multi-head cross-attention and a \ac{FFN} module. The masked multi-head self-attention takes output of the previous block (summation of the previous character embedding and positional encoding if the self-attention in the first block) as query, key and value with a mask which filter out future information. The multi-head cross-attention receives output of the self-attention as query and output of encoder as key and value. The output of the multi-head cross-attention pass through the \ac{FFN} and then is sent to the next block. \textcolor{red}{CE} 

Due to the hybrid training framework, the total loss function is formulated as the weighted sum of \ac{CTC} loss and attention-based loss, formulated as:
\begin{equation}\label{eq:loss_function}
    \mathcal{L} = \lambda p_{ctc}(Y|R) + (1-\lambda) p_{att}(Y|R)
\end{equation}
where $p_{ctc}$ and $p_{att}$ are losses of \ac{CTC} and attention, $\lambda$ is the coefficient to control influences of two loss components.

\subsection{Predict-and-Update vs. feature concatenation}
The proposed predict-and-update strategy is different from feature concatenation in terms of the design concept and the actual implementation. First, with predict-and-update, we use visual signals to predict the character posterior that is phonetically informing. The fusion takes place between the phonetically informed output distributions and the raw audio features. However, in feature concatenation, the raw features of audio and visual signals are simply put together in the early stage without reference to a text or phonetic content. {Second, the visual prediction is achieved through several Conformer blocks that effectively use a video sequence ahead of the arrival of audio sounds, similar to that in the visual cueing mechanism  \cite{schwartz2004seeing}, to prime the listener when and on which frequency to focus.}

\section{Experiment}\label{sec:Experiment}
In the visual cueing mechanism by humans, % In this paper, we study the humans speech perception of that preceding video modality cues listeners when and on which frequency of speeches to focus. We presume there are two distinct characteristics in this manner: 
we observe two properties, 1) a preceding visual signal primes the listener what sounds are expected, 2) the expected sounds serve as the cue to the audio signal. %, as temporal-frequency audio information is a kind of audio raw features. 
The P\&U net is designed to study the effect of the above two properties. %  by this manner, we would like to validate this manner's superiority and disentangle the contribution of cueing mechanism and cueing raw audio features (early cueing) in this manner.
We first design experiments where the visual embedding primes the update encoder at different stages of audio sequence modelling, \ie early cueing vs. late cueing. We then compare the proposed methods with some \ac{STOA} systems. As a contrastive system, we also evaluate the traditional 
feature concatenation technique in comparison with the cueing mechanism. We also perform ablation studies to observe the contributions of the individual modules. All experiments are conducted on large \ac{AVSR} data sets consisting of uncontrolled spoken sentences.
% As cognition studies show evidence of early fusion in human speech perception and we argue the importance of early fusion, we first apply the factorized \ac{FFN} on different cross-modal Conformer blocks to support our arguments. These experiments are conducted on large \ac{AVSR} dataset which containing uncontrolled natural language sentences to make conclusion more convincing. We also provide analysis of why fusion happens in earlier stage achieves better performance. Moreover, comparison between our proposed method and some \ac{STOA} is presented to further illustrate advantages of early fusion.

\subsection{Dataset}
\textbf{LRS2-BBC} \cite{afouras2018deep}: The dataset is an audio-visual collection of more than 143K utterances with a 60K vocabulary from BBC programs. All samples are carefully pre-processed by face detection, shot detection, face tracking, facial landmark detection, audio-visual synchronization, forced audio-subtitle alignment, and alignment verification. The facial images are captured  from both profile and frontal views. The dataset is split into 4 parts: pre-train, train, val, and test. {The pre-train set and the train set are for training, while the val set and the test set are for development and test, respectively.} %\textcolor{red}{(in machine learning, we typically have train, validation, and test sets. The validation is used to fine-tune parameters, but evaluation set = test set. i am confused here.)} \textcolor{burntumber}{The 4 parts are split by the dataset provider. In pre-train set, the provider also provide force alignment label. And the reason they offer val and test is because they follow the manner of some AI competitions. In these competitions, the host only offers val set and keep the test set secretly, and the final performance depends on the test set to consider model ability of generalization.}
The pre-train set comes with a word-level force alignment. The utterances in the other three parts are shorter than 6 seconds, while a small portion of the utterances in the pre-train set are longer than 6 seconds. The training data is around 224 hours.

\textbf{LRS3-TED} \cite{afouras2018deep}: The dataset contains 164K utterances and have a similar vocabulary size to LRS2-BBC, but with longer utterances. The dataset has a total of 438 hours and is split into 3 parts: pre-train, train, and test. The pre-processing of LRS3-TED is the same as that of LRS2-BBC. Like in LRS2-BBC, the pre-train set also comes with word-level force alignment. However, unlike LRS2-BBC, LRS3-TED has no speaker overlap between the training and test sets, because the data of LRS3-TED is from TED videos and is split by presenters.

We choose character-level tokens as ground truth since words in two datasets are unconstrained. There are 40 output classes, namely $C=40$, including the 26 characters (A-Z), the 10 digits (0-9), the token [space] for separating words, [prime] ('), [sos] indicates the start or end of an utterance, and [blank] for \ac{CTC} training.

\subsection{Data augmentation}

In addition, we inject babble noise into waveforms before \ac{STFT} and implement a method similar to SpecAugment \cite{park2019specaugment} after \ac{STFT} to augment the data. In detail, we apply two masks each of which is up to 0.4 seconds on the time axis while we randomly mask two frequency bands where each band is less than 1k Hz. The maximum time warp is 5 frames. We generate the babble noise by 
mixing samples presented in \cite{afouras2018deep}. During training, we follow the \ac{SNR} distribution of \cite{ma2021end}, particularly, an uniform distribution over [No noise, 20dB, 15dB, 10dB, 5dB, 0dB, -5dB].  

For video images, patches of 112$\times$112 are horizontally flipped with 50\% random selection.
% further cropped randomly with 50\% horizontal flipping as inputs.

%\subsection{Evaluation Metrics}
%Following some \ac{STOA} methods \cite{ma2021end,afouras2018deep}, we employ \ac{WER} to evaluate the proposed method and compare it with other \ac{STOA} methods. The \ac{WER} is defined as

%\begin{equation}\label{eq:wer}
 % WER =  \frac{S+D+I}{N}
%\end{equation}
%where $S$, $D$ and $I$ indicates the number of substitutions, deletion and insertions between ground truth and decoded hypothesis by edit distance algorithm \cite{wagner1974string}. $N$ denotes the number of words in the ground truth.

\subsection{Language Model}

Following \cite{ma2021end}, we adopt a transformer-based \cite{irie2019language} character-level \ac{LM}. The training corpus has 16.2 million words, which consists of the transcriptions of LibriSpeech (960 hours) \cite{panayotov2015librispeech}, pre-train and train sets of LRS2-BBC \cite{afouras2018deep} and LRS3-TED \cite{afouras2018deep}. Each character is encoded to a 128-D vector without positional encoding. $d_a$ and $h$ of the \ac{MHSA} and $d_f$ of the \ac{FFN} in transformer blocks are 8, 512 and 2048, respectively. Besides, the \ac{LM} contains 16 transformer blocks. We train the \ac{LM} 
by Adam optimizer \cite{kingma2014adam} for 30 epochs with the val set of LRS2-BBC as the development set, learning rate $10^{-4}$, and batch size 32. \ac{LM} training is implemented by ESPnet \cite{watanabe2018espnet} with a single GeForce GTX 1080 Ti (11 GB memory).

\subsection{Pre-training}
\label{pretrain_setting}
% The prediction part is to predict distribution sequences towards text. In other words, we hope the output of the prediction encoder is highly correlated with text. Thus, we pre-train the prediction encoder by a lipreading task. The network architecture follows visual Conformer \cite{ma2021end} whose visual encoder and visual frontend are the same as our method shown in \fig~\ref{fig:architecture}. 
As described in \Sec~\ref{subsec:prediction}, we pre-train the predictor via a lipreading task that takes visual frame sequence as input and generates character posteriors.  
In the predictor, there are 12 Conformer blocks in the visual Conformer encoder, namely $N_p=12$. The hyperparameters are $d_k=256$, $d_a=256$, $d_{ff}=2048$, $h=4$. And the kernel size of the depthwise 1D convolutional layers is 31. Besides, for LRS2-BBC, the visual frontend is trained, which is the same as \fig~\ref{fig:architecture}. But for LRS3-TED, due to limited disk storage and lengthy training procedure if we train the frontend simultaneously, we adopt parameters of the visual frontend from \cite{afouras2018deep} and pre-process image sequences to 512-D visual embeddings via the frozen visual frontend. Because of pre-processing, there is no visual data augmentation for LRS3-TED and we just crop central patches of 112$\times$112 to the visual frontend in the pre-processing. This setting is consistent in the following \ac{AVSR} experiments.

%Besides, we design the factorized-excitation \ac{FFN} as it does not change architectures of the \ac{FFN} parameters. Therefore, it's potential to benefit from pre-trained \ac{ASR} models with the \ac{FFN}.
% which is a great advantage of representation fusion and decision fusion.
{In the update encoder, we adopt the factorized-excitation \ac{FFN} instead of a standard \ac{FFN}, as shown in \fig~\ref{subfig:ccconformer}. As both share the same architecture, we can pre-train an audio Conformer \cite{ma2021end} to initialize the update encoder}. The audio Conformer serves as an audio-based ASR model but replaces its input filter-bank features or convolutional features by spectrum from \ac{STFT}. There are 12 Conformer blocks in the \ac{ASR} Conformer encoder. The hyper-parameters of the \ac{ASR} encoder are the same as that of the lipreading encoder except for the attention head $h=8$.

The pre-trained lipreading and \ac{ASR} model share the same decoder architecture as that of the P\&U net in \Sec~\ref{subsec:loss}, which consists of 6 transformer blocks. The hyper-parameters of \ac{MHSA}, \ac{FFN} of the decoder block are set as
$d_k=256$, $d_a=256$, $d_{ff}=2048$, $h=8$.
% the same as the counterparts in the \ac{ASR} encoder.

\begin{table}[!htb]
\centering
\caption{Performance of video-only model given the metric of \ac{CER}. LM: Language model.
}\label{tab:pretrained}
\begin{tabular}{c|ccc}
\toprule \midrule
\textbf{Inference Method}& \ac{CTC} & \ac{CTC}\&LM & \ac{CTC}\&Decoder\&LM \\
\midrule
\textbf{LRS2-BBC} & 32.2\% & 28.8\% & 27.2\%\\
\textbf{LRS3-TED} & 40.1\% & 36.5\% & 34.6\%\\
\midrule \bottomrule
\end{tabular}
\end{table}

% \begin{table}[!htb]
% \centering
% \caption{Performance of audio-only and video-only model. Metric is \ac{WER}. The noisy means original utterances with 0 dB babble noise.
% }\label{tab:pretrained}
% \begin{tabular}{c|ccc}
% \toprule \midrule
% \textbf{Modality}& Audio-only (clean) & Audio-only (noisy) & Video-only \\
% \midrule

% \textbf{LRS2-BBC} & 4.3\% & 28.8\% & 40.2\%\\
% \textbf{LRS3-TED} & 3.3\% & 43.7\% & 52.0\%\\
% \midrule \bottomrule
% \end{tabular}
% \end{table}

We report the performance of the pre-trained lipreading model in \tab~\ref{tab:pretrained}. The \ac{CTC} inference results are solely based on the output of the predictor $\rho$. The \ac{CTC}\&LM denotes the system that infers a text sequence by incorporating an external language model of characters. The \ac{CTC}\&Decoder\&LM system further involves the decoder. 
When the \ac{CTC}, the decoder, and the \ac{LM} are all involved during frame-by-frame inference, %the \ac{LM} predicts a probability distribution of the next character based on character frames which have been predicted, the decoder decodes a probability distribution of the next character based on existed character frames and the visual context representation, the \ac{CTC} module calculates a probability distribution that the \ac{CTC} inference is in accord with existed character frames and different next characters. 
the output of \ac{LM}, the decoder, and the \ac{CTC} module are summed up with their respective weight of $\psi$, $\gamma$, and $1-\gamma$ to form a final probability distribution. The top-1 character is used as the output. We empirically set $\gamma=0.1$, $\psi=0.6$ for LRS2-BBC. As the \ac{LM} is optimized on the val set
of LRS2-BBC, there could be a mismatch with LRS3-TED. Therefore, we choose $\gamma=0.2$, $\psi=0.4$ for LRS3-TED. Moreover, beam search is employed with a width of 20.

It is reported that the \ac{CTC} inference by the predictor alone achieves a character error rate of 32.2\% and 40.1\% for top-1 decoding on LRS2-BBC and LRS3-TED datasets. By incorporating the \ac{LM} and decoder, the lipreading model is further improved. As we use the visual embedding $\rho$ rather than the top-1 decoding as the visual cue, we believe that the lipreading model provides informative visual cues.

%\textcolor{red}{However, \ac{CTC} inference is based on frame-independent assumption and can not perfectly reflect recognition ability of the predictor. By incorporating the \ac{LM} and the decoder, performances further increase. All these results shows that the predictor is capable of roughly predicting target sequences.}

\subsection{Implementation details}
\label{subsec:details}

The predictor and the update encoder are initialized by the parameters of the encoders of the pre-trained lipreading and \ac{ASR} models, respectively, while the decoder parameters are initialized by the decoder of the lipreading model. For the variant of the visual embedding $\rho'$, its dimension $K=32$ because 32 is the closest number to the number of output classes and meanwhile is the divisor of the hidden vector dimension of vanilla \ac{FFN} $d_{ff}$. Consequently, the dimension of audio sub-spaces $d_l=64$.
%$K=32$ and
% The update encoder contains 12 cross-modal Conformer blocks. However, some of the blocks may be substituted by the vanilla Conformer block, because of ablation studies in \ref{subsec:early_fusion}. The hyper-parameters of the update encoder is same with that of the visual encoder except $h=8$.

We implement our models on a single GeForce RTX 3090 (24 GB memory). The network is trained by Adam optimizer \cite{kingma2014adam} with $\beta_1=0.9$, $\beta_2=0.98$, and $\epsilon=10^{-9}$. Batch sizes are 8 for LRS2-BBC and 16 for LRS3-TED. Optimizers take a step every 4 and 2 batches for two dataset. The learning rate linearly increases to $2\times10^{-4}$ by 25000 steps and afterwards decreases proportionally to the inverse square root of the step number, which is employed in \cite{vaswani2017attention}. 

For experiments on LRS2-BBC, we mix its pre-train and train sets as the training set and exclude samples with more than 24 seconds. Because the mixed training set contains more than 140K utterances, we process them by virtual epochs, randomly picking 16,384 samples from the mixed set. The network is trained for 500 virtual epochs. And only samples within 6 seconds will be picked at the first 100 epochs, which is similar to curriculum learning \cite{afouras2018deep}.

For experiments on LRS3-TED, the virtual epoch, the mixture of pre-train and trainval set, and the curriculum learning are employed as well. The difference is that we only exclude samples of more than 48 seconds, and we train the network for 750 virtual epochs, as LRS3-TED are bigger than LRS2-BBC. 

At the inference phase, the attention decoder module, the \ac{CTC} module and the \ac{LM} are employed together. We set the same weights as that in \ref{pretrain_setting}, namely, $\gamma=0.1$, $\psi=0.6$ for LRS2-BBC, $\gamma=0.2$, $\psi=0.4$ for LRS3-TED. %\textcolor{red}{to Jiadong: i don't understand why val is the evaluation set, and it becomes less effective for LRS3-TED.}
%\textcolor{burntumber}{Development set may be more accurate here. We train the language model to minimize perplexity on the val of the LRS2-BBC. LRS2-BBC is from British BBC programms but the LRS3-TED from TED talks whose talkers are from all over the world. The text in two set may be not in the same domain.}
% As the \ac{LM} is optimized on the val set
% of LRS2-BBC, there could be a mismatch with LRS3-TED. Therefore, we choose $\gamma=0.2$, $\psi=0.4$ for LRS3-TED. Moreover, beam search is employed with width 20.

% Besides, we implement a representative early fusion method (Feat Concat) to compare it with the P\&U net (early4) to demonstrate the benefit of cueing mechanism. The Feat Concat fuses audio and visual features described in \ref{audio_feat} and \ref{visual_feat} by concatenation. Before concatenation, two features are normalized to a similar scale. After concatenation, the concatenated audio-visual feature reduces to a 256-D feature by passing through two linear layers and then is fed to 12 Conformer blocks ($d_k=256$, $d_a=256$, $d_{ff}=2048$, $h=8$). The loss function of the Feat Concat is same with the P\&U net (early4).
Besides, we implement the feature concatenation (Feat Concat) as a contrastive model of the P\&U net (early4) to show the benefit of cueing mechanism. The Feat Concat fuses audio and visual features described in \Sec~\ref{audio_feat} by simply joining the time-aligned audio and visual features into a single feature. Before  the concatenation, two features are normalized to a similar scale. Subsequently, the concatenated audio-visual feature vector is reduced to a 256-D vector by passing through two linear layers, that is then taken by the 12 Conformer blocks ($d_k=256$, $d_a=256$, $d_{ff}=2048$, $h=8$). The loss function of the Feat Concat is the same as the P\&U net.

\subsection{Cueing at different stage}
\label{subsec:early_fusion}
%We check the contribution of early cueing to human speech perception by experiments where the visual prediction cues different level of audio features during sequence modelling. 
As shown in \fig~\ref{subfig:punetearly}-\ref{subfig:punetlate}, the order of the cross-modal Conformer blocks and the Conformer block is inter-changeable. There are a total of 12 blocks in the update encoder. In particular, we design an experiment with five configurations: P\&U net (early4) and P\&U net (early8)
as \fig~\ref{subfig:punetearly}: First 4 and 8 blocks are cross-modal Conformer blocks; P\&U net (middle): middle 4 blocks are cross-modal Conformer blocks; P\&U net (late) as \fig~\ref{subfig:punetlate}: last 4 blocks are cross-modal Conformer blocks; and P\&U net (all): All 12 blocks are cross-modal Conformer blocks. We report the results in the \tab~\ref{tab:different_level}, where ``clean" and ``noisy" denote the original test samples without or with 0 dB babble noise.

\begin{table}[!htb]
\centering
\caption{Performance comparison in \ac{WER} of visual embedding cues different levels of audio features  with external \ac{LM}. Noisy: under 0 dB babble noise.
}\label{tab:different_level}
\begin{tabular}{ccccc}
\toprule \midrule

\textbf{Dataset} & \multicolumn{2}{c}{\textbf{LRS2-BBC}}  & \multicolumn{2}{c}{\textbf{LRS3-TED}} \\

% \textbf{Dataset} & \textbf{LRS2-BBC} & & \textbf{LRS3-TED} & \\
\midrule
  & \textbf{Clean} & \textbf{Noisy} & \textbf{Clean} & \textbf{Noisy} \\ \midrule

P\&U net (early4) & \textbf{3.83}\% & \textbf{9.50}\% & \textbf{3.05}\% & \textbf{8.48}\%\\
P\&U net (middle) & 4.11\% & 11.33\% & 3.14\% & 9.69\%\\
P\&U net (late) & 4.61\% & 14.23\% & 3.62\% & 12.75\%\\
\midrule
P\&U net (early8) & \textbf{3.71}\% & \textbf{9.89}\% & \textbf{3.26}\% & \textbf{8.51}\%\\
P\&U net (all) & 3.87\% & 11.62\% &  3.52\%&  9.78\%\\
\midrule \bottomrule
\end{tabular}
\end{table}

%0.28\% absolute,  0.5\% absolute, a 0.09\% absolute, a 0.48\% absolute,

Under the clean condition, the results of LRS2-BBC in \tab~\ref{tab:different_level} show that the P\&U net (early4) achieves a \ac{WER} of 3.83\%, which outperforms the P\&U net (middle), whose \ac{WER} is 4.11\% with a relative \ac{WER} reduction of 6.8\%.
% From the results on clean LRS2-BBC in the \tab~\ref{tab:different_level}, we observe that the performance of  P\&U net (early4) is 3.83\% \ac{WER}, while that of P\&U net (middle) is 4.11In other words, P\&U net (early4) reduces 6.8\% relative \ac{WER}. 
The P\&U net (late) only achieves a \ac{WER} of 4.61\%, which lags behind the P\&U net (middle) by a 10.8\% relative \ac{WER}.
% There is a 10.9\% relative \ac{WER} reduction between P\&U net (middle) and (late), namely 4.11\% vs 4.61\%.
These results suggest that fusion at an earlier stage is more effective than fusion at a later stage as far as visual cueing is concerned. % that the later the visual prediction supplies, the more deteriorated performance is obtained. % (6.8\% to 10.9\%).
%Besides, there is a trend of accelerated \ac{WER} deterioration with a later cueing \ie~from 6.8\% to 10.9\%.  
% These results suggest the importance of early cueing that the visual prediction attending to earlier blocks of the update encoder yields better performance.
% These results also suggest that the performance may deteriorate quickly with late fusion (6.8\% to 10.9\%). 
The same trend is observed on LRS3-TED. 
The P\&U net (early4), P\&U net (middle) and P\&U net (late) achieve the \ac{WER} of 3.05\%, 3.14\%, and 3.62\%. % with the incremental relative \ac{WER} equals 2.9\% and 13.3\%, respectively. 

We use the decoding of an utterance as an example in \tab~\ref{tab:example_LRS2} to show how the different systems behave. The visual cue with an early cueing provides an accurate phonetic prediction that cannot be perceived in the audio signal, \ie the non-audible release of /d/ at the end of ``PICKLED".
 
\begin{table}[!htb]
\centering
\caption{An \ac{AVSR} example in the clean environment. Audio: Pre-trained \ac{ASR} model; Video: Pre-trained lipreading model; GT: Ground Truth.
}\label{tab:example_LRS2}
\begin{tabular}{cc}
\toprule \midrule
 & \textbf{Transcription}   \\ \midrule
\textbf{GT} & THAT'S A PICKLED WALNUT  \\
\textbf{P\&U net (early4)} & THAT'S A PICKLED WALNUT   \\
\textbf{P\&U net (middle)} & THAT'S A PICKLE WALNUT   \\
\textbf{P\&U net (late)} & THAT TO PICKLE WALNUT   \\
\textbf{Audio} & THAT'S A PICKLE WALNUT   \\
\textbf{Video} & THAT'S THE BUILDING \\
% \textbf{All blocks} & THAT'S A PICKLE WALNUT & 25  \\
\midrule \bottomrule
\end{tabular}
\end{table}
 
%  there are 6.8\% and 10.9\% relative \ac{WER} reductions between P\&U net (early4) and P\&U net (middle)
Under the noisy condition, speech signals are corrupted by babble noise with the resulting \ac{SNR} of 0 dB. We observe the same performance trend as under the clean condition. On LRS2-BBC, the P\&U net (middle) only achieves a \ac{WER} of 11.33\%, which is worse than the P\&U net (early4), whose \ac{WER} is 9.5\%, by a relative reduction of 16.2\%. 
The P\&U net (late) obtains a \ac{WER} of 14.23\%, which lags behind the P\&U net (middle) by a relative reduction of 20.4\%. The results in LRS3-TED are consistent with those in LRS2-BBC.
The relative \ac{WER} reductions from the P\&U net (early4) to (middle), and from the P\&U net (middle) to (late) are 12.5\% (9.69\% to 8.48\%) and 24\% (12.75\% to 9.69\%), respectively.
% The trend of accelerated deterioration also exists under the noisy environment.
Moreover, we observe that 
the performance drops from P\&U net (middle) to (late) is greater than that from P\&U net (early4) to (middle), \ie 20.4\% vs. 16.2\% on LRS2-BBC, and 24\% vs. 12.5\% on LRS3-TED. All results strongly suggest applying visual cueing at an earlier stage is beneficial.

%\textcolor{blue}{Beside similarities in the clean and noisy conditions, there are differences where the performance drops above}
% respectively. Although the trends under clean and noisy conditions are the same, the quantities of deterioration under noisy environment are
%are higher than the counterparts \ie~10.9\% and 6.8\% under clean environment.
% The relative \ac{WER} reduction between the P\&U net (early4) and the P\&U net (middle) under noisy and clean conditions are 6.8\% and 16.2\%.
% We also observe a larger performance gap than that in the clean environment.
% This phenomenon suggests
The results also show that early cueing is more effective in a noisy environment than late cueing. We can observe that the performance drops from P\&U net (early4) to (middle) are 6.8\% and 16.2\% under the clean and noisy conditions on LRS2-BBC, respectively. The same phenomenon exists in such a comparison on LRS3-TED.
%This results confirm the benefit of early cueing in the noisy environment.

\begin{table}[!htb]
\centering
\caption{An example of AV result in the noisy environment. Audio: Pre-trained \ac{ASR} model; Video: Pre-trained lipreading model; GT: Ground Truth; P\&U net is omitted before (early4), (middle) and (late).
}\label{tab:example_noisy_LRS2}
\begin{tabular}{p{0.7cm}<{\centering}p{6.1cm}<{\centering}}
\toprule \midrule
 & \textbf{Transcription}   \\ \midrule
\textbf{GT} & INSTEAD OF BEING SOMEONE'S ARM CANDY \\
\textbf{Early4} & INSTEAD OF BEING SOMEONE'S ARM CANDY   \\
\textbf{Middle} & INSTEAD OF BEING SOMEONE'S FAMILY   \\
\textbf{Late} & INSTEAD OF BEING SOMEONE'S AUNTIE   \\
\textbf{Audio} & INSTEAD OF BEING SOMEONE'S ON KENT  \\
\textbf{Video} & SELF BECAUSE IT WASN'T COMPLICATED  \\
% \textbf{All blocks} & THAT'S A PICKLE WALNUT & 25  \\
\midrule \bottomrule
\end{tabular}
\end{table}

We use the decoding of an utterance as an example in \tab~\ref{tab:example_noisy_LRS2} to show how the different systems behave under 0 dB babble noise. The visual cue with an earlier fusion provides an accurate phonetic prediction that differentiates /m/ and /n/ in a better way than the audio cue alone.

%\tab~\ref{tab:example_noisy_LRS2} shows an example under 0 dB babble noise that P\&U net (early4) is able to differentiate the acoustically less distinguishable pair of /m/ and /n/ while others can not. The \ac{ASR} model wrongly recognizes the syllable /m/ in the 'ARM' as /n/ and therefore interprets 'ARM' as 'on'. The P\&U net (late) can not correctly distinguish /m/ as well. But the P\&U net (middle) distinguishes /m/ although it fails to recognize "CANDY". At last, the P\&U net (early4) successes to transcript the speech by earlier cueing.

%  there are 6.8\% and 16.2\% relative \ac{WER} reductions compared with P\&U net (early4) and P\&U net (middle) in the clean and noisy environment, respectively. While these differences are 10.9\% and 20.4\% between P\&U net (middle) and P\&U net (late). From these figures of comparison, we can see a trend that the visual prediction attending to an earlier block of the update encoder yield a better performance. Meanwhile, earlier fusion boosts \ac{AVSR} more in the noisy environment than clean environment.  \textcolor{red}{more analysis}
 
% It is necessary to validate the importance of early fusion on the LRS3-TED because of its distinct characteristics: 1. There is no speaker overlap between training data and test test; 2. It is larger data than LRS2-BBC (approximating 2 times). From \tab~\ref{tab:different_level},

In \tab~\ref{tab:different_level}, we also observe that the P\&U net (early8) hardly outperforms the P\&U net (early4). Furthermore, the P\&U net (early4) exceeds the P\&U net (all) across the board, suggesting that additional later cross-modal cueing is not contributing. %(to Jiadong: this is a problematic results, it is theoretically wrong. We shouldn't elaborate. \textcolor{burntumber}{To Prof. Li, need we add sentences about our hypothesis of this phenomenon like following?}\textcolor{blue}{What may account for this phenomenon is that features cued by the visual embedding in additional later blocks may be similarly or more representative of characters with the visual embedding.}
%\textcolor{red}{(to Jiadong: this is against the logic, (all) should be as good as (early4) in theory. it cannot be worse mathematically.)} \textcolor{burntumber}{To Prof. Li, I think the reason of the worse performance is that the features in the last 4 block of P\&U net (all) have been more representative about the text than the visual prediction, so it may be a disturbance they are cued by the visual prediction, which cause the worse performance. I think it is the best when the visual prediction only cue features which are less representative than the visual prediction.}
%, which indirectly proves the importance of early cueing. 
 
% To further dig out underlying reasons why earlier fusion performs better in the noisy environment, we check the robustness of $R$ in noisy environment with P\&U net (early4), P\&U net (middle) and P\&U net (late).    

% \begin{table*}[!htb]
% \centering
% \caption{Similarities between audio-visual representation $R$ in the clean environment and other noise condition.
% }\label{tab:noise-invariant}
% \begin{tabular}{c|ccccccc}
% \toprule \midrule
% \textbf{Attention block}  & \textbf{20 dB} & \textbf{15 dB} & \textbf{10 dB} & \textbf{5 dB} & \textbf{0 dB}  & \textbf{-5 dB} & \textbf{-10 dB}\\ \midrule

% Early 4 blocks & 0.990 & 0.980 & 0.964 & 0.938 & 0.900 & 0.865  & 0.861\\
% Middle 4 blocks & 0.992 & 0.985 & 0.971 & 0.950 & 0.918 & 0.891 & 0.902\\
% Late 4 blocks & 0.994 & 0.987 & 0.973 & 0.949 & 0.919 & 0.923 & 0.959\\
% baseline & 0.987 & 0.972 & 0.94.5 & 0.885 & 0.778 & 0.721 & 0.824\\

% \midrule \bottomrule
% \end{tabular}
% \end{table*}

\subsection{Comparative study}
\label{subsec:comparative_study}

We compare the P\&U net and other \ac{STOA} systems on LRS2-BBC in \tab~\ref{tab:stoa_LRS2}. 
% In this table, P\&U net indicates our proposed method which only adopts the cross-modal Conformer block at the first 4 blocks of the update encoder and employ the vanilla Conformer block as other parts of the update encoder.
We observe that the P\&U net (early4) significantly outperforms other systems both under clean and noisy conditions. Particularly, let us use the best-performing AV Conformer \cite{ma2021end} (fusion of separate audio and visual context representation) as the baseline, which has the same learnable parameter as ours. The P\&U net (early4) achieves a relative \ac{WER} reduction of 16.3\%, \ie from a 4.3\% \ac{WER} to a 3.6\% \ac{WER}, over the baseline under the clean condition. We also observe a 42.0\% relative \ac{WER} reduction, \ie from a 15.7\% \ac{WER} to a 9.1\% \ac{WER}, under the noisy condition.%.5\% absolute (11.6\% relative) \ac{WER} and 6.2\% absolute (39.5\% relative)
 %This improvement validates the superiority of human speech perception both in clean and noisy environment.
% And it also indirectly supports that humans recognize speeches by an audio-visual early fusion.   

\begin{table}[!htb]
\centering
\caption{Comparison of our proposed method with some \ac{STOA} methods on the LRS2-BBC dataset. Entries in columns of clean and noisy indicate \ac{WER} with external \ac{LM}. Noisy: original waveform plus 0 dB babble noise.
}\label{tab:stoa_LRS2}
\begin{tabular}{p{2.3cm}<{\centering}p{3.5cm}<{\centering}p{0.7cm}<{\centering} p{0.6cm}<{\centering}}
\toprule \midrule
\textbf{Method}  & \textbf{Training data} & \textbf{Clean} & \textbf{Noisy} \\ \midrule

TM-seq2seq\cite{afouras2018deep} & MVLRS(730)+LRS2\&3$^{0.4}$(632) & 8.5\% & 34.2\% \\
TM-CTC\cite{afouras2018deep} & MVLRS(730)+LRS2\&3$^{0.4}$(632) & 8.2\% & 23.6\% \\
DCM\cite{lee2020audio} & LRS2(224) & 8.6\% & 28.9\% \\
CTC/Attention\cite{petridis2018audio} & LRW(157)+LRS2(224) & 7.0\% & - \\
TDNN\cite{yu2020audio} & LRS2(224) & 5.9\% & - \\
AV Conformer*\cite{ma2021end} & LRS2(224) & 4.3\% & 15.7\% \\
\midrule
Feat Concat  & LRS2(224) & 4.4\% & 11.5\% \\
P\&U net (late) & LRS2(224) & 4.6\% & 14.2\% \\
P\&U net (early4) & LRS2(224) & \textbf{3.6}\% & \textbf{9.1}\% \\
\midrule\bottomrule
\end{tabular}
\end{table}

\footnote{*we re-implement the algorithm using inputs, noise, training procedure, \ac{LM} described in this paper.}

To appreciate the contribution of the visual cueing mechanism, %, we also want to validate merits of the cueing mechanism.
% In addition to taking advantage of early cueing, the P\&U net (early4) also inherits the merits of the cueing mechanism.
% And the predict-and-update manner which we is good at relieving the influence of noise on a modality or a sensory that is only involved in the update step. In other words, this manner should strengthen P\&U in the noisy environment, which may accounts for why outperformance of the P\&U net (early4) in the noisy environment is about 2.5 times than that in clean environment (42.0\% vs 16.3\%).However, because the P\&U net (early4) is the combination of the cueing mechanism and early cueing, it is hard to clarify that the great outperformance of the P\&U net (early4) to the baseline benefits from only the either or the both. 
we first compare the worst-performing P\&U net (late) and the baseline. The P\&U net (late) conducts audio-visual fusion after audio features have passed through 8 Conformer blocks and its performance with clean utterance is worse than that of the baseline (4.6\% vs. 4.3\%). But the P\&U net (late) still outperforms the baseline by a 9.6\% relative \ac{WER} (14.2\% vs. 15.7\%) in the noisy environment, as shown in \tab~\ref{tab:stoa_LRS2}. We are convinced that the visual cueing mechanism contributes to the performance gain by P\&U net under the noisy condition. 

We also compare the P\&U net (early4) and the Feat Concat. Both of them adopt an early fusion strategy, but differ in how audio-visual signals are fused, as discussed in \Sec~III-E. As shown in \tab~\ref{tab:stoa_LRS2}, the P\&U net (early4) outperforms the Feat Concat under both clean (3.6\% \ac{WER} vs. 4.4\% \ac{WER}) and noisy condition (9.1\% \ac{WER} vs. 11.5\% \ac{WER}) by approximately 20\% relative \ac{WER} reduction. We attribute the performance gain to the visual cueing mechanism.

We further compare the systems on LRS3-TED in \tab~\ref{tab:stoa_LRS3}. We  observe that the P\&U net (early4) outperforms other \ac{STOA} methods, especially under the noisy condition.
In particular, the P\&U net (early4) leads AV Conformer by 10\% relative \ac{WER} (3.0\% vs. 3.3\%) and 42.4\% relative \ac{WER} (8.3\% vs. 14.4\%) under the clean and noisy condition, respectively. Besides, the P\&U net (early4) also outperforms Feat Concat under both the clean and noisy conditions, which confirms the benefit of cueing mechanism again.

\begin{table}[!htb]
\centering
\caption{Comparison of our proposed method with some state-of-the-art methods on the LRS3-TED. Entries in clean and noisy columns indicate \ac{WER} with external \ac{LM}. Noisy: original waveform plus 0 dB babble noise.
}\label{tab:stoa_LRS3}
\begin{tabular}{cccc}
\toprule \midrule
\textbf{Method}  & \textbf{Training data} & \textbf{Clean} & \textbf{Noisy} \\ \midrule

TM-seq2seq\cite{afouras2018deep} & MVLRS(730)+LRS2\&3$^{0.4}$(632) & 7.2\% & 42.5\% \\
TM-CTC\cite{afouras2018deep} & MVLRS(730)+LRS2\&3$^{0.4}$(632) & 7.5\% & 27.7\% \\
DCM\cite{lee2020audio} & LRS3$^{0.4}$(438) & 8.8\% & 30.9\% \\
EG-s2s\cite{xu2020discriminative} & LRS3$^{0.0}$(474) & 6.8\% & 25.5\% \\
RNN-T\cite{makino2019recurrent} & YT(31k) & 4.5\% & - \\
Attentive Fusion\cite{wei2020attentive} & LRS3$^{0.4}$(438) & 6.4\% & - \\
AV Conformer*\cite{ma2021end} & LRS3$^{0.4}$(438) & 3.3\% & 14.4\% \\
\midrule
Feat Concat  & LRS3$^{0.4}$(438) & 3.3\% & 10.1\% \\
P\&U net (late) & LRS3$^{0.4}$(438) & 3.6\% & 12.8\%\\
P\&U net (early4) & LRS3$^{0.4}$(438) & \textbf{3.0}\% & \textbf{8.3}\% \\
\midrule \bottomrule
\end{tabular}
\end{table}

\subsection{Visual cueing vs. context representation concatenation}

To appreciate how the visual cueing mechanism works under noisy condition, we compare the audio-visual context representation $R$, i.e. the output of the update encoder in the P\&U net (late), and the fusion of audio and visual context representations in the baseline. We measure the cosine similarity as expressed in Eq.~\ref{eq:cossim} between the audio-visual context representations $R$ for speech signals without noise and $R'$ for that with different \ac{SNR} noise. 

\begin{equation}\label{eq:cossim}
  \theta^j_t =  \frac{R^j_t\cdot R'^{j}_t}{||R^j_t||\cdot||R'^{j}_t||}
\end{equation}
where $j$ and $t$ indicate an utterance index and a frame index, respectively. %Specifically, $R$ denotes the audio-visual context representations given clean inputs while $R'$ is under noisy inputs. 
$\theta \in [-1, 1]$ denotes the cosine similarity. 

We report the average cosine similarities of the two networks across all time frames and utterances on the LRS2-BBC test set in the \tab~\ref{tab:noise-invariant}. The cosine similarity $\theta$ of both networks approaches 1.0 when \ac{SNR} $\geq 15$ dB.
We also observe that the cosine similarity of the P\&U net (late) is consistently higher than that of the baseline. As the \ac{SNR} decreases, $\theta$ of the baseline deteriorates rapidly, while $\theta$ of the P\&U net (late) remains steady.  
The results in \tab~\ref{tab:noise-invariant} account for why the P\&U net (late) performs worse than the baseline under the clean condition but otherwise under 0 dB \ac{SNR}, which shows the benefit of the visual cueing mechanism under the noisy condition.

\begin{table}[!htb]
\centering
\caption{The cosine similarity between the audio-visual context representations $R$ and $R'$ for speech signals at different \ac{SNR} on the LRS2-BBC dataset.
}\label{tab:noise-invariant}
\setlength{\tabcolsep}{1.6mm}{
\begin{tabular}{c|ccccccc}
\toprule \midrule
\textbf{Methods}  & \textbf{20 dB} & \textbf{15 dB} & \textbf{10 dB} & \textbf{5 dB} & \textbf{0 dB}  & \textbf{-5 dB} \\ \midrule

P\&U net (late) & \textbf{0.994} & \textbf{0.987} & \textbf{0.973} & \textbf{0.949} & \textbf{0.919} & \textbf{0.923} \\
AV Conformer*\cite{ma2021end} & 0.987 & 0.972 & 0.94.5 & 0.885 & 0.778 & 0.721 \\

\midrule \bottomrule
\end{tabular}
}
\end{table}

\subsection{Initialization by pre-trained models}
\begin{figure}[]
    \centering
    \includegraphics[scale=0.35]{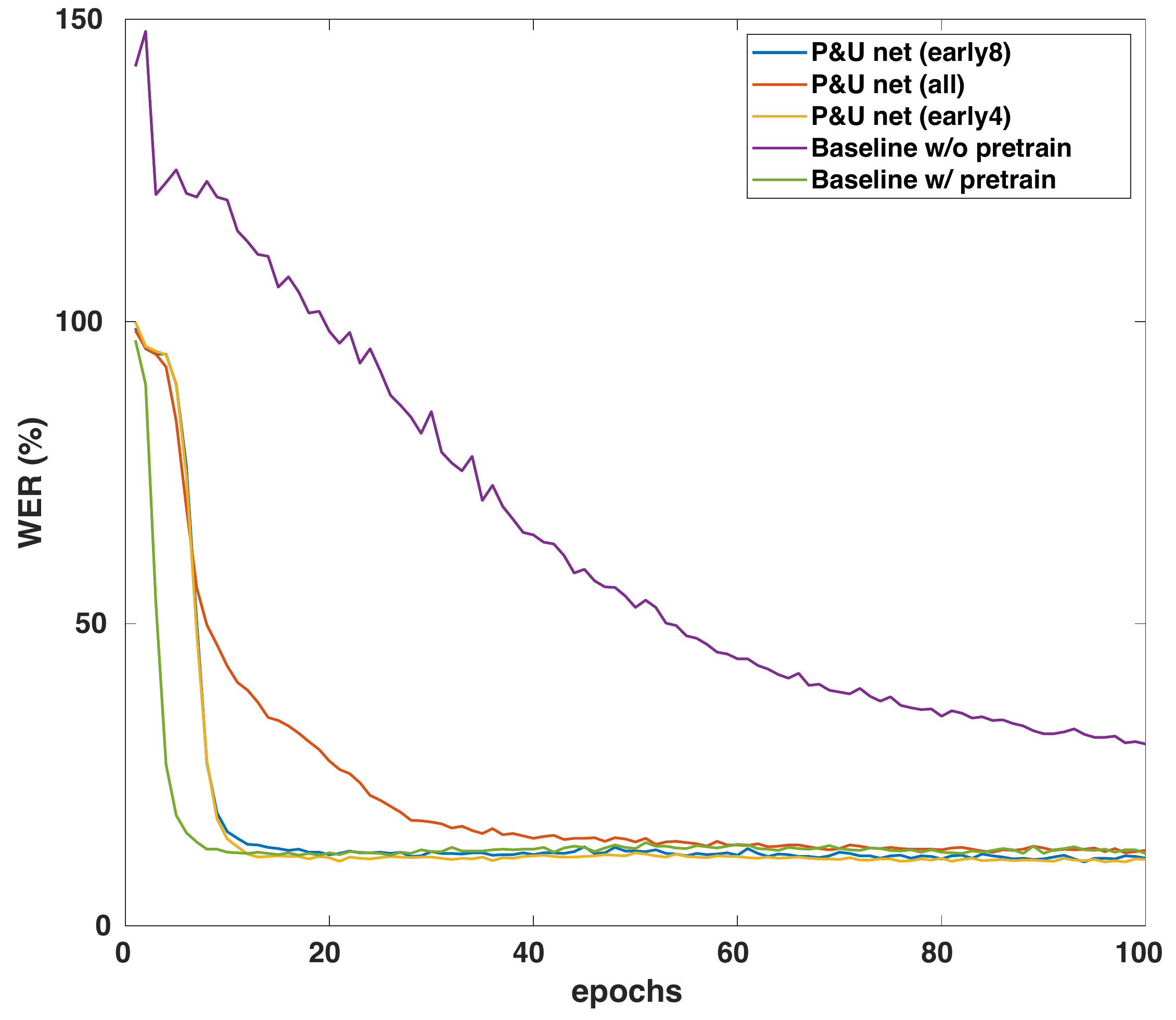}
    \caption{The \ac{WER} training curves on LRS2-BBC val set in a teacher forcing mode.}
    \label{fig:training_speed}
\end{figure}

The P\&U net  benefits from the pre-trained \ac{ASR} and lipreading models as described in \Sec~\ref{pretrain_setting}. \fig~\ref{fig:training_speed} shows the \ac{WER} as a function of the number of training epochs for the P\&U net and the baseline AV Conformer. The P\&U nets are initialized by the pre-trained unimodal models. We also compare AV Conformer with random initialization as a contrast. We can see that the AV Conformer converges more quickly with the pre-trained models than without. %\ac{WER}s of our methods, especially P\&U net (early4) and P\&U net (early8), also drop quickly, which validates the significance
% the potential of our proposed early fusion 
%of adopting pre-trained unimodal models. 
%Therefore, it is potential to use the models pretrained on larger dataset for P\&U net initialization to yield better performance although we do not employ extra data to pre-train our network.
% may use a unimodal system trained on a larger dataset to initialize our P\&U net to yield better performance. 

\subsection{Dimension of visual embedding}
We are interested in the effect of the pre-defined parameters of the factorized-excitation \ac{FFN}, for example, the number of audio feature subspaces $K$, which also determines the dimension of audio feature sub-spaces $d_l$. %dimension \ie, K to see the influence.
% Since our method mimics the manner that video modality cues audio features and $K$ influences the dimensions of the visual prediction, it is desirable to explore the influence of variation of $K$.
% We also ablate, which depends on $K$.
% Although we have clarified the reason of $C=32$ in \sec~\ref{subsec:details},
To this end, we design an experiment on LRS2-BBC and LRS3-TED with four different $K$=16, 32, 64, 128, thus, $d_l$=128, 64, 32, 16
% The experiment is conducted on
with the P\&U net (early4), and report the WER results in \tab~\ref{tab:vp_dimension}.

\begin{table}[!htb]
\centering
\caption{Experiments on the visual embedding, \ie character posteriors, for clean speech signals and noisy speech signals corrupted by 0 dB babble noise.
}\label{tab:vp_dimension}
\begin{tabular}{ccccc}
\toprule \midrule
\textbf{Dataset} & \multicolumn{2}{c}{\textbf{LRS2-BBC}}  & \multicolumn{2}{c}{\textbf{LRS3-TED}}  \\ \midrule
\textbf{Dimension}  &  \textbf{Clean} & \textbf{Noisy} &  \textbf{Clean} & \textbf{Noisy}\\ \midrule

\textbf{K=16} &  3.63\% & \textbf{9.13}\% &  3.02\% & 8.25\%\\
\textbf{K=32} & 3.83\% & 9.50\% &  3.05\% & 8.48\%\\
\textbf{K=64} & 3.67\% & 9.97\% &  3.07\% & \textbf{8.15}\%\\
\textbf{K=128} & \textbf{3.62}\% & 9.16\% &  \textbf{3.01}\% & 8.38\%\\
\midrule \bottomrule
\end{tabular}
\end{table}

It is observed that the dimension $K$ does not significantly affect the results under both clean and noisy conditions and in both LRS2-BBC and LRS3-TED datasets. Therefore, in the rest of the experiments, we have chosen $K$=32.  %, , on LRS2-BBC, the P\&U nets with the largest $K$=128 and the smallest $K$=16 perform similarly and better than the rests. Meanwhile, even the worst performance under clean environment with $K$=32 achieves a \ac{WER} of 3.83\%, which lags behind the best performance with $K$=128 by only an absolute \ac{WER} reduction of 0.21\%. The gap between the best and worst performance under noisy environment is an absolute \ac{WER} of 0.84\% (9.97\% with $K$=64 vs 9.13\% with $K$=16). These gaps further reduce on the experiment on LRS3-TED. The gap with clean inputs is an absolute \ac{WER} reduction of 0.06\% (3.07\% with $K$=64 vs 3.01\% with $K$=128). While the gap under noisy environment is an absolute \ac{WER} reduction of 0.33\% (8.48\% with $K$=32 vs 8.15\% with $K$=64). Therefore, we can conclude that variation of $K$ does not significantly influence cueing effect of video modality and there is no extra effort required to choose $K$.

% \begin{table}[!htb]
% \centering
% \caption{Comparison of dimension of the visual prediction. Noisy: 0 dB babble noise.
% }\label{tab:vp_dimension}
% \begin{tabular}{ccccc}
% \toprule \midrule
% & LRS2-BBC & & LRS3-TED & \\ \midrule
% \textbf{Dimension}  &  \textbf{clean} & \textbf{Noisy} &  \textbf{clean} & \textbf{Noisy}\\ \midrule

% 16 &  3.6\% & 9.1\% &  3.0\% & 8.3\%\\
% 32 & 3.8\% & 9.5\% &  3.2\% & 8.3\%\\
% 64 & 3.7\% & 10.0\% &  3.0\% & 8.3\%\\
% 128 & 3.6\% & 9.2\% &  3.3\% & 8.6\%\\
% \midrule \bottomrule
% \end{tabular}
% \end{table

\subsection{Position of the factorized-excitation \ac{FFN}}

We design the cross-modal Conformer block by replacing an \ac{FFN} module with a factorized-excitation \ac{FFN}. 
As shown in \fig~\ref{subfig:conformer},  there are two \ac{FFN} modules in a Conformer block. We alter the position of the factorized-excitation \ac{FFN} by
 1) replacing the first \ac{FFN} module , 2) replacing the second \ac{FFN} module  and 3) replacing both \ac{FFN} modules. All three configurations are implemented on  the P\&U net (early4) with the audio subspace dimension $d_l=64$.

\begin{table}[!htb]
\centering
\caption{Experiments on the position of the factorized-excitation (FE) \ac{FFN} for clean speech signals and noisy speech signals corrupted by 0 dB babble noise.
}\label{tab:fffn_position}
\begin{tabular}{ccc}
\toprule \midrule
\textbf{Position of FE FFN}  &  \textbf{Clean} & \textbf{Noisy} \\ \midrule
\textbf{first \ac{FFN}} &  \textbf{3.66}\% & \textbf{9.28}\% \\
\textbf{second \ac{FFN}} & 3.83\% & 9.50\% \\
\textbf{both \ac{FFN}} & 4.25\% & 10.86\% \\
\midrule \bottomrule
\end{tabular}
\end{table}

As  in \tab~\ref{tab:fffn_position} we observe that replacing the first \ac{FFN} performs slightly better than the second one, which confirms the benefit of early cueing. Replacing both \ac{FFN} does not achieve performance gain suggests that visual cue plays a secondary role in the audio-visual speech recognition. %performs worse than the others by at least 10\% relative \ac{WER}. This might because vision affects audio too much, considering the unsatisfactory performance of the video-only system reported in \tab \ref{tab:pretrained}.
% The reason may because  the visual prediction as a cue influences audio features too much, but it is not very accurate as shown in \tab~ \ref{tab:pretrained}.  

\section{Conclusion}

We have proposed a novel end-to-end audio-visual speech recognition network architecture, \ie the P\&U net. This study is motivated by the finding in human speech perception where visual signals prime the listener before the arrival of audio signals. We hypothesize that the visual cueing mechanism and early fusion are the two contributing factors to effective audio-visual speech recognition. The experiments have validated our hypotheses. The fact that the visual cueing mechanism has a clear advantage over the simple fusion suggests that the interaction between audio and visual signals is as important as the signals themselves as far as speech recognition is concerned.

\ifCLASSOPTIONcaptionsoff
  \newpage
\fi



% http://www.michaelshell.org/tex/bibtex/
%\bibliographystyle{IEEEtran}

%   0.5em minus 0.4em\mathbb{R}lax Harlow, England: Addison-Wesley, 1999.

% \end{thebibliography}
\bibliographystyle{IEEEtran}
\bibliography{NUSref2}

\end{document}